\documentclass[
preprint,
nofootinbib,
amsmath,amssymb,
aps,
floatfix,
]{revtex4-1}

\usepackage{graphicx}
\usepackage{dcolumn}
\usepackage{bm}
\usepackage{hyperref}
\usepackage{color}
\usepackage{epstopdf}
\usepackage[utf8]{inputenc}
\usepackage{morefloats}

\begin{document}
\allowdisplaybreaks

\title{\Large Gradient resummation for nonlinear chiral transport: an insight from holography}
\author{Yanyan Bu}\email{yybu@hit.edu.cn}
\affiliation{Department of Physics, Harbin Institute of Technology, Harbin 150001, China}

\author{Tuna Demircik}\email{demircik@post.bgu.ac.il}
\author{Michael Lublinsky}\email{lublinm@bgu.ac.il}
\affiliation{Department of Physics, Ben-Gurion University of the Negev, Beer-Sheva 84105, Israel}


\begin{abstract}
Nonlinear transport phenomena induced by  chiral anomaly are explored within a 4D field theory defined holographically as
$U(1)_V\times U(1)_A$ Maxwell-Chern-Simons theory in Schwarzschild-$AdS_5$. In presence of weak constant background electromagnetic fields,
the constitutive relations for vector and axial currents,  resummed to all orders in the gradients of charge densities, are encoded in nine momenta-dependent
transport coefficient functions (TCFs). These TCFs are first calculated analytically up to third order in gradient expansion, and then  evaluated numerically beyond the hydrodynamic limit.
Fourier transformed, the TCFs become memory functions. The memory function of the chiral magnetic effect (CME) is  found to differ dramatically  from the instantaneous response form
of the original CME. Beyond hydrodynamic limit and when external magnetic field is larger than some critical value,
the chiral magnetic wave (CMW) is discovered to possess  a discrete spectrum of non-dissipative modes.

\end{abstract}

\date{\today}

\maketitle
\tableofcontents

\section{Introduction}\label{intro}

In this paper we continue exploring  hydrodynamic regime of  relativistic plasma with chiral asymmetries. We closely follow previous works \cite{1608.08595,1609.09054}  focusing on massless fermion plasma with two Maxwell gauge fields, $U(1)_V\times U(1)_A$.
Dynamics of hydrodynamic theories is governed by conservation equations (continuity equations) of the currents.
As a result  of  chiral anomaly,  which appears in relativistic QFTs with massless fermions,
global $U(1)_A$ current  coupled to external electromagnetic (e/m) fields is no longer conserved.
The continuity equations turn into
\begin{equation}\label{cont eqn}
\partial_{\mu}J^{\mu}=0, \qquad \qquad \partial_{\mu}J_5^{\mu}=12 \kappa \vec{E}\cdot \vec{B},
\end{equation}
where $J^{\mu}, J_5^{\mu}$ are vector and axial currents and $\kappa$ is an anomaly coefficient ($\kappa=eN_c/(24\pi^2)$ for $SU(N_c)$ gauge theory with a massless Dirac fermion in fundamental representation and $e$ is  electric charge, which will be set to unit from now on). $\vec{E}$ and $\vec{B}$ are external vector electromagnetic fields.

The continuity equations could be regarded as time evolution equations for the charge densities $\rho$ ($\rho_5$)  sourced by three-current $\vec J$ ($\vec J_5$).
However, these equations cannot be solved as an initial value problem without additional input, the currents $\vec J$ and $\vec J_5$.  In hydrodynamics,
the currents are expressed in terms of thermodynamical variables, such as the charge densities $\rho$ and $\rho_5$ themselves, temperature $T$,
and the  external e/m fields $\vec{E}$ and $\vec{B}$ if present.
These are known as constitutive relations, which generically take the form
\begin{equation} \label{const}
\vec J\ =\ \vec J\;[\rho, \rho_5, T, \vec E, \vec B]; \hspace{3cm}  \vec J_5=\vec J_5[\rho, \rho_5, T, \vec E, \vec B].
\end{equation}
The constitutive relations should be considered as  ``off-shell'' relations, because they treat the charge
density $\rho$ ($\rho_5$) as independent of $\vec J$  ($\vec J_5$). Once (\ref{cont eqn}) is imposed, the currents' constitutive relations (\ref{const}) are put ``on-shell''.

In addition to the charge current sector discussed above, one  has to simultaneously consider energy-momentum conservation. In general, these two dynamical sectors are coupled.  However, in the discussion below, we will ignore  back-reaction of the charge current sector on the energy-momentum conservation. This will be referred to as {\it probe limit}.

In the long wavelength limit,  the constitutive relations are usually presented as a (truncated) gradient expansion.
At any given order, the gradient expansion is fixed by thermodynamic considerations and symmetries,
up to a finite number of transport coefficients (TCs).
The latter should be  either computed from the underlying microscopic theory  or deduced experimentally. Diffusion constant,
DC conductivity or shear viscosity are examples of the lowest order TCs.

It is well known, however,  that in relativistic theory  truncation of the gradient expansion at any fixed order
leads to serious conceptual problems
such as violation of causality.
Beyond conceptual issues, causality violation results in numerical instabilities rendering the entire framework unreliable.
Causality is restored  when all order gradient terms are included, in a way providing a UV completion to the ``old'' hydrodynamic effective theory.
Below we will refer to such case as \textit{all order resummed} hydrodynamics \cite{Lublinsky:2007mm,0905.4069,1406.7222,1409.3095,1502.08044,1504.01370}. The first completion of the type
was originally proposed by  M\"{u}ller, Israel, and Stewart  \cite{Muller1967,ISRAEL1976310,ISRAEL1976213,ISRAEL1979341} who  introduced retardation effects in the constitutive relations for the currents. Formulation of \cite{Muller1967,ISRAEL1976310,ISRAEL1976213,ISRAEL1979341}
is the most popular scheme  employed  in practical simulations.
Essentially, all order resummed hydrodynamics is equivalent to a non-local constitutive relation of the type (here we take the charge diffusion current as an example):
\begin{equation} \label{diff memory}
\vec J_{\rm diff}(t)=\int_{-\infty}^{+\infty}dt^\prime \tilde{\mathcal{D}}(t-t^\prime) \vec\nabla \rho(t^\prime),
\end{equation}
where $\tilde{\mathcal{D}}$ is the memory function of the diffusion function $\mathcal{D}(\omega,q^2)$ \cite{1511.08789}, which is generally non-local both in time and space.
Causality implies that $\tilde{\mathcal{D}}(t)$ has no support for $t <0$. In practice, the memory function is typically modelled: M\"{u}ller-Israel-Stewart formulation \cite{Muller1967,ISRAEL1976310,ISRAEL1976213,ISRAEL1979341} models the memory functions with a simple exponential in time parametrised by a
relaxation time.

Chiral plasma plays a major role in a number of fundamental research areas, historically starting from primordial plasma in the
early universe \cite{Kuzmin:1985mm,Vilenkin:1982pn,Rubakov:1996vz,Grasso:2000wj,doi:10.1142/S0218271804004530}. During the last decade,
macroscopic effects induced by the chiral anomaly were found to be of relevance in relativistic heavy ion collisions \cite{025420,Kharzeev:2013ffa,Huang:2015oca}, and have been searched intensively at RHIC and LHC \cite{1512.05739,1610.00263,1708.01602,1708.08901,Skokov:2016yrj}.
Finally, (pseudo-)relativistic systems in condensed matter physics, such as Dirac and
Weyl semimetals, display anomaly-induced phenomena, which were recently observed experimentally \cite{Liu:2014,Lv:2015pya,Xu:2015cga,2014ARCMP,1412.6543,1503.01304,1507.06470} and can be studied via similar theoretical methods \cite{1207.5808,Landsteiner:2014vua,Jimenez-Alba:2015awa,Landsteiner:2015lsa}.

The constitutive relations (\ref{const}) are well known to receive contributions induced by the chiral anomaly.
The most familiar example is the \textit{chiral magnetic effect} (CME) \cite{PhysRevD.22.3080,0808.3382,Fukushima:2010vw}:  a vector current is generated along an external magnetic field  when  a chiral imbalance between left- and right-handed fermions is present ($\vec J\sim \rho_5\vec B$). Another important transport phenomenon induced by the chiral anomaly is the \textit{chiral separation effect} (CSE) \cite{Son:2004tq,Metlitski:2005pr}:
left and right charges get separated along an applied external magnetic field ($\vec J_5\sim \rho\vec B$).
Combined,  CME and CSE lead to a  new gapless excitation called \textit{chiral magnetic wave} (CMW) \cite{Kharzeev:2010gd}. This is a propagating wave along the magnetic field. There is a vast literature on CME/CSE and other chiral anomaly-induced transport phenomena, which we cannot review here in full. We refer the reader to recent reviews \cite{1207.5808,Kharzeev:2013ffa,Kharzeev:2015znc,Huang:2015oca,Kharzeev:2015kna} and references therein on the subject of chiral anomaly-induced transport phenomena.

Beyond naive CME/CSE, there are (infinitely) many additional  effects induced or affected by chiral anomaly. Particularly, transport phenomena
{\it nonlinear} in external fields were  realised recently \cite{Avdoshkin:2014gpa} to be of critical importance in having a self-consistent evolution of chiral plasma. This argument, together with the causality discussions mentioned earlier, would lead to the conclusion that the constitutive relations (\ref{const}) should contain infinitely many ``nonlinear'' transport coefficients in order to guarantee  applicability of the constitutive relations in a  broader regime. Recently, this triggered strong interest  in nonlinear chiral transport phenomena within chiral kinetic theory (CKT)  \cite{1603.03620,1603.03442,1705.01267,1710.00278}.
Previous works on the subject of nonlinear chiral transport phenomena
include \cite{1105.6360} based on the notion of entropy current, and \cite{1304.5529} based on the fluid-gravity correspondence \cite{0712.2456}.

The objective of present work is to explore  all order gradient resummation for nonlinear transport effects induced by the chiral anomaly\footnote{The asymptotic nature  of the gradient expansion and problems related to  resummation of the series have been a hot topic over the last few years,
see recent works \cite{1302.0697,1503.07514,1509.05046,1707.02282}.
In our approach, however, we never attempt to actually sum the series and thus these discussions are of no relevance to our formalism.}, further extending the results of Refs. \cite{1608.08595,1609.09054,Bu:2018psl}.

Just like in  Refs. \cite{1608.08595,1609.09054,Bu:2018psl},
our playground will be a holographic model, that is
 $U(1)_V\times U(1)_A$ Maxwell-Chern-Simons theory in Schwarzschild-$AdS_5$ \cite{Yee:2009vw,Gynther:2010ed}
to be introduced in Section \ref{s2}, for which we know to compute a zoo of transport coefficients   exactly.
Hoping for some sort of universality, we could learn from this model about both general phenomena and  relative strengths of various effects.

In our recent publication  \cite{Bu:2018psl}, we reviewed all different studies which were performed in \cite{1608.08595,1609.09054,Bu:2018psl}.
Those studies  and the present one are largely independent even though performed within the same holographic model. For brevity,
we will not repeat this review here, but will make  connection to these previous works whenever relevant. We refer reader to \cite{Bu:2018psl} for the summary of the different approximations which employed in these series of works and the present work. The comparison of the resultant constitutive  relations and our comments about the total current is also presented there.

Anomalous transport phenomenon is frequently discussed from the viewpoint of its dissipative nature and, equivalently, its  contribution to entropy production \cite{0906.5044, 1105.6360,Neiman:2010zi,Lin:2011mr,Bhattacharya:2012zx}.  CME is well known to be non-dissipative \cite{1207.5808,Kharzeev:2013ffa,1502.01547}.  What about
the dissipative nature of other anomalous transport phenomena, say beyond CME?
In  \cite{1105.6360} the  transport coefficients that are odd in $\kappa$  were identified as anomaly-induced and, based on space parity $\mathcal{P}$ arguments,
are claimed to be non-dissipative. This is to distinguish from anomaly-induced corrections to normal transports, which appear to be even in $\kappa$.  While the $\mathcal{P}$-based arguments  seem to work perfectly for the second order hydrodynamics \cite{1105.6360}, a more natural criterion of dissipation seems to be based on  time-reversal symmetry $\mathcal{T}$.  $\mathcal{T}$-odd transport coefficients describe dissipative currents, whereas $\mathcal{T}$-even ones are non-dissipative \cite{1105.6360}. The anomaly-induced phenomena explored  below will involve terms both dissipative and not.

In the next Section, we will review our results including  connections to the previous works \cite{1608.08595,1609.09054,Bu:2018psl}. The following Sections present details
of  the calculations.

\section{Summary of the results}\label{Summary}

The objective of \cite{1608.08595,1609.09054,Bu:2018psl} and of the present  work  is to systematically explore (\ref{const}) under different approximations.
Following  \cite{1608.08595,1609.09054,Bu:2018psl}, the charge densities  are split into constant backgrounds and space-time dependent fluctuations
\begin{equation} \label{linsch1}
\begin{split}
&\rho(x_\alpha)=\bar{\rho}+\epsilon\delta\rho(x_\alpha), \qquad\qquad \rho_5(x_\alpha)=\bar{\rho}_5+\epsilon\delta\rho_5(x_\alpha),\\
&\vec{E}(x_\alpha)=\vec{\bf{E}}+\epsilon\delta\vec{E}(x_\alpha), \qquad\quad \vec{B}(x_\alpha)=\vec{\bf{B}}+\epsilon\delta\vec{B}(x_\alpha),
\end{split}
\end{equation}
where $\bar{\rho}$, $\bar{\rho}_5$,  $\vec {\bf E}$ and $\vec {\bf B}$ are the constant backgrounds, while
$\delta\rho$, $\delta\rho_5$,  $\delta\vec{E}$, $\delta\vec{B}$ stand for the fluctuations. Here
$\epsilon$ is a formal expansion parameter to be used below.
Furthermore, being most of the time unable to perform calculations for arbitrary
background fields, we introduce  an expansion in the field strengths
\begin{equation} \label{linsch2}
\vec{\bf{E}}\rightarrow\alpha\vec{\bf{E}}, \quad\quad \vec{\bf{B}}\rightarrow\alpha\vec{\bf{B}},
\end{equation}
where $\alpha$ is the corresponding  expansion parameter.
Below we will introduce yet another expansion parameter $\lambda$,
which will correspond to a gradient expansion. For the purpose of the gradient counting, e/m fields will be  considered as  $\mathcal{O}(\lambda^1)$.

Throughout this work, the e/m backgrounds $\vec {\bf E}$ and $\vec {\bf B}$ are treated as weak. The constitutive relations (\ref{const}) can be formally expanded both  in $\epsilon$ and $\alpha$
\begin{equation}\label{Jnn}
\begin{split}
&J^t=\rho,~~~~~~~~~~~~~~~\vec{J}=\vec{J}^{\;(0)(1)}+ \vec{J}^{\;(1)(0)}+ \vec{J}^{\;(1)(1)}+\cdots,\\ &J^t_5=\rho_5,~~~~~~~~~~~~~~\vec{J}_5=\vec{J}^{\;(0)(1)}_5 + \vec{J}^{\;(1)(0)}_5+ \vec{J}^{\;(1)(1)}_5 +\cdots,
\end{split}
\end{equation}
where the first superscript denotes order in $\epsilon$ and the second in $\alpha$. $\vec{J}^{\;(0)(1)}$, $\vec{J}^{(1)(0)}$, $\vec{J}_5^{\;(0)(1)}$ and $\vec{J}_5^{\;(1)(0)}$ were derived in \cite{1608.08595}.

The goal of present paper is to extend the work initiated in \cite{1608.08595} by
computing $\vec{J}^{\;(1)(1)}$ and $\vec{J}^{\;(1)(1)}_5$. Particularly, we will evaluate transport coefficients functions (TCFs) associated with relevant nonlinear transport phenomena discovered in \cite{Bu:2018psl} via a fixed order gradient expansion. For simplification, we turn off the fluctuations of the external e/m fields, $\delta\vec E=\delta\vec B=0$. At $\mathcal{O}\left(\epsilon^1 \alpha^1\right)$, the currents take the following  forms
\begin{eqnarray} \label{ji11}
\vec{J}^{\;(1)(1)}&&= \sigma_{\bar{\chi}} \kappa \vec{\bf{B}} \delta\rho_5- \frac{1}{4} \mathcal{D}_H  (\bar{\rho}\vec{\bf{B}}\times\vec{\nabla}\delta\rho)- \frac{1}{4} \bar{\mathcal{D}}_H  (\bar{\rho}_5\vec{\bf{B}}\times\vec{\nabla}\delta\rho_5) -\frac{1}{2}\sigma_{a\chi H} (\vec{\bf{E}}\times\vec{\nabla}\delta\rho_5) \nonumber\\
&&- \frac{1}{2}\bar{\sigma}_{a\chi H} (\vec{\bf{E}}\times\vec{\nabla}\delta\rho) +\sigma_1 \kappa \left[(\vec{\bf{B}}\times\vec{\nabla})\times\vec{\nabla}\right] \delta\rho
+ \sigma_2 \kappa \left[(\vec{\bf{B}}\times\vec{\nabla})\times\vec{\nabla}\right] \delta\rho_5 \nonumber \\
&&+ \sigma_3 \kappa \left[(\vec{\bf{E}}\times\vec{\nabla})\times\vec{\nabla}\right] \delta\rho + \bar\sigma_3 \kappa\left[(\vec{\bf{E}}\times\vec{\nabla})\times\vec{\nabla} \right] \delta\rho_5,
\end{eqnarray}
\begin{eqnarray} \label{j5i11}
\vec{J}_5^{\;(1)(1)}&&=\sigma_{\bar{\chi}} \kappa \vec{\bf{B}} \delta\rho-\frac{1}{4} \mathcal{D}_H (\bar{\rho}\vec{\bf{B}}\times\vec{\nabla}\delta\rho_5) - \frac{1}{4} \bar{\mathcal{D}}_H (\bar\rho_5\vec{\bf{B}}\times\vec{\nabla} \delta\rho) - \frac{1}{2}\sigma_{a\chi H} (\vec{\bf{E}}\times\vec{\nabla} \delta\rho)\nonumber \\
&& - \frac{1}{2}\bar\sigma_{a\chi H} (\vec{\bf{E}}\times\vec{\nabla}\delta\rho_5) +\sigma_1 \kappa \left[(\vec{\bf{B}}\times\vec{\nabla})\times\vec{\nabla}\right] \delta\rho_5 + \sigma_2 \kappa \left[(\vec{\bf{B}}\times\vec{\nabla})\times\vec{\nabla} \right] \delta\rho\nonumber \\
&&+ \sigma_3 \kappa \left[(\vec{\bf{E}}\times\vec{\nabla})\times\vec{\nabla}\right] \delta\rho_5 + \bar\sigma_3 \kappa\left[(\vec{\bf{E}}\times\vec{\nabla})\times\vec{\nabla}\right] \delta\rho ,
\end{eqnarray}
where all the coefficients are scalar functionals of the derivative operator $\partial_\mu$
\begin{align} \label{sigmas op}
&\sigma_{\bar \chi}[\partial_t, \vec \nabla],~\mathcal{D}_H[\partial_t, \vec \nabla],~\bar{\mathcal{D}}_H[\partial_t, \vec \nabla],~\sigma_{a \chi H}[\partial_t, \vec \nabla],~\bar\sigma_{a \chi H}[\partial_t, \vec \nabla],~ \sigma_{1,2,3}[\partial_t, \vec \nabla],~ \bar\sigma_{3}[\partial_t, \vec \nabla].
\end{align}
Thanks to the linearisation, the constitutive relations (\ref{ji11}, \ref{j5i11}) could be conveniently presented in Fourier space. Then, the functionals (\ref{sigmas op}) are turned into functions of frequency and spatial momentum, $(\partial_t, \vec \nabla)\to(-i\omega, i\vec{q}) $, which
we refer to as TCFs \cite{1409.3095}. TCFs contain information about infinitely many derivatives and associated transport coefficients. In practice, they are not computed as a series resummation of order-by-order  hydrodynamic expansion, and are in fact exact to all orders.  TCFs go beyond the hydrodynamic low frequency/momentum limit and contain collective effects of non-hydrodynamic modes. Fourier transformed back into real space, TCFs turn into memory functions, cf. (\ref{diff memory}).

Except for the $\bar{\sigma}_{a\chi H}$-term, all the rest of the terms in (\ref{ji11}, \ref{j5i11}) have already appeared in our previous publication \cite{Bu:2018psl} at a fixed order in the gradient expansion. The novelty of present study is to {\it consistently} generalise many of the TCs of \cite{Bu:2018psl} into TCFs, guaranteeing applicability of the constitutive relations (\ref{ji11}, \ref{j5i11}) in a broader regime.


To the best of our knowledge, the TCF $\sigma_{\bar{\chi}}$ is introduced here for the first time and will play a crucial role below, see (\ref{CMW}).
It is important to stress the difference between $\sigma_{\bar{\chi}}$ and $\sigma_\chi$ of \cite{1608.08595,Yee:2009vw,Landsteiner:2013aba}. Both TCFs generalise CME/CSE. Yet, while the latter is induced by spacetime variation
of the magnetic field, the former is due to inhomogeneity of the charge densities $\rho,\rho_5$. One might naively expect that both TCFs are equal. In fact they are not, as we demonstrate below.
For comparison, here we quote the hydrodynamic expansion of the CME TCF $\sigma_\chi$ which was calculated in \cite{1608.08595}
\begin{equation}
\begin{split}\label{sigmachi}
\sigma_\chi&=6\left\{1+ i\omega \log 2 -\frac{1}{4} \omega^2 \log^22 -\frac{q^2}{24} \left[\pi^2-432\kappa^2\left(\bar{\rho}^2_{_5}+ 3\bar{\rho}^2 \right)\left(\log 2-1\right)^2\right]\right\}+\cdots.
\end{split}
\end{equation}
As seen  from (\ref{sigmachi}, \ref{sigmab}) the first order gradient corrections to CME/CSE (i.e., the relaxation time corrections) are different depending on if it is the magnetic field or the charge density that varies with time. In addition, while $\sigma_\chi$ depends on $\rho,\rho_5$ nonlinearly, $\sigma_{\bar\chi}$ does not depend on $\rho,\rho_5$ at all.

The TCF $\sigma_{\bar{\chi}}$ enters the dispersion relation of CMW:
\begin{equation}\label{CMW}
\omega=\pm \sigma_{\bar{\chi}}(\omega,q^2) \,\kappa\vec q\cdot\vec {\bf B} - i \mathcal{D}(\omega,q^2)  q^2,
\end{equation}
which is exact to all orders in $q^2$. In the hydro limit, using (\ref{D}, \ref{sigmab}), the dispersion relation can be solved analytically with the most comprehensive result reported in \cite{Bu:2018psl}.  Yet, we have discovered a set of solutions with purely real $\omega$ in the present work.
That is, for some (continuum set of) values of magnetic field $\bf B$, there is a discrete density wave mode $(\omega_B, q_B)$, which propagates without any dissipation (Figure \ref{ab}).
This is a quite intriguing result, which originates solely from the all order resummation procedure. The details about the non-dissipative discrete density wave mode are deferred to subsection \ref{s44}.

\begin{figure}[htbp]
\centering
\includegraphics[width=0.5\textwidth]{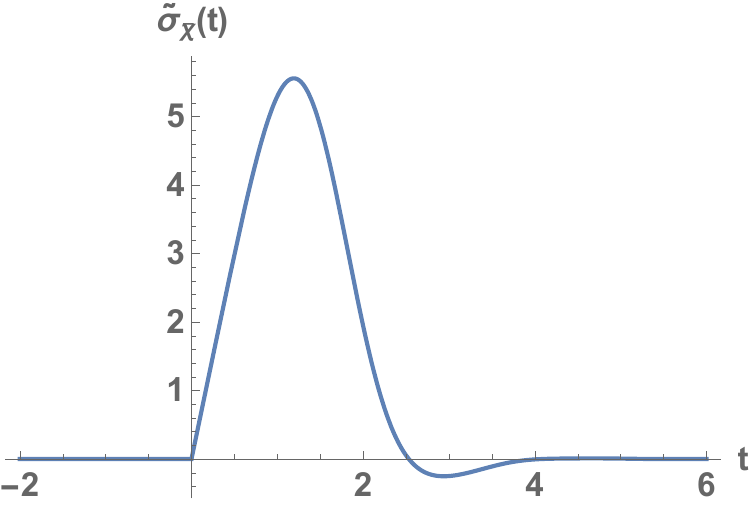}
\caption{The memory function $\tilde{\sigma}_{\bar{\chi}}(t)$ when $q=0$.}
\label{memory}
\end{figure}

As mentioned in the Introduction, TCFs could be Fourier transformed into memory functions, for an extensive discussion see e.g. \cite{1511.08789,1502.08044}. The CME current with retardation effects is
\begin{equation}
\vec J_{\rm CME}(t) =\kappa\,{\vec{\bf B}} \int_{-\infty}^{\infty}dt^\prime\,\tilde{\sigma}_{\bar{\chi}}(t-t^\prime)\delta \rho_5(t^\prime)
\end{equation}
Via inverse Fourier transform, the CME/CSE memory function is  (we focus on  the case $q=0$),
\begin{equation} \label{sigmab_t}
\tilde{\sigma}_{\bar{\chi}}(t)\equiv \frac{1}{\sqrt{2\pi}} \int_{-\infty}^{+\infty} d\omega e^{-i\omega t} \sigma_{\bar\chi}(\omega,q=0).
\end{equation}
The memory function  $\tilde \sigma_{\bar\chi}$ is displayed in Figure \ref{memory}.  An important feature of this function is that it has no support at negative times, which is nothing but manifestation of causality.  Another very interesting observation is that rather than having an instantaneous response picked at the origin, like in original CME, the actual response
is significantly delayed and picked at a finite time of order temperature.  This behaviour of $\tilde \sigma_{\bar\chi}$  is quite distinct from
diffusion memory function $\tilde{\mathcal{D}}(t)$ and shear viscosity memory functions computed previously in \cite{1511.08789,1502.08044}, which are  picked at the origin.

Let us briefly comment on the remaining terms. $\mathcal{D}_H$ generalises the Hall diffusion $\mathcal{D}_H^0$ \cite{1603.03442,1603.03620,Bu:2018psl} into a TCF of $\omega,q^2$; $\bar{\mathcal{D}}_H$ is just its axial analogue. So, we will refer to $\mathcal{D}_H$ and $\bar{\mathcal{D}}_H$ as Hall diffusion functions. $\sigma_{a\chi H}$ is a TCF extending the anomalous chiral Hall conductivity $\sigma_{a\chi H}^0$ \cite{1603.03442,1603.03620,Bu:2018psl}.
$\bar{\sigma}_{a\chi H}$ could be considered as an axial analogue of $\sigma_{a\chi H}$. However, as will be clear later, $\bar{\sigma}_{a\chi H}$ has an overall factor $q^2$ so that it will be non-vanishing starting from fourth order in the  gradient expansion only.

$\sigma_{1,2,3}$ and $\bar{\sigma}_3$ are TCFs of  the third order derivative operators (we remind the reader that the e/m fields are counted as of first order).
$\sigma_1,\sigma_2$ correspond to rotor of Hall diffusion \cite{Bu:2018psl}, and $\sigma_3,\bar{\sigma}_3$ are rotors of anomalous chiral Hall effect \cite{Bu:2018psl}.

Each TCF in (\ref{ji11}, \ref{j5i11}) can be split into real part (even powers of frequency) and imaginary part (odd powers of frequency). Based on the time reversal criterion,
we conclude that the real parts of $\sigma_{\bar{\chi},1,2}$, $\mathcal{D}_H$, $\bar{\mathcal{D}}_H$ and imaginary parts of $\sigma_{a\chi H},\bar{\sigma}_{a\chi H}$,$ \sigma_3, \bar\sigma_3$ are non-dissipative; all the rest do lead to dissipation of the currents. It is interesting to notice that there are points in the ($\omega,q$) phase space, where some of the dissipative terms vanish. Particularly,  this happens to $\rm Re[\mathcal{D}]$ and $\rm Im[\sigma_{\bar{\chi}}]$. This feature leads to presence of a non-dissipative discrete density wave mode which is mentioned earlier.

The constitutive relations (\ref{ji11}, \ref{j5i11}) could be re-written in a more compact way,
\begin{equation}
\begin{split}
J^{(1)(1)}_i=&\sigma_{\bar{\chi}}\kappa {\bf B}_i \delta\rho_5- \kappa{\bf B}_i \left(\sigma_1\vec{\nabla}^2\rho + \sigma_2\vec{\nabla}^2\rho_5 \right) -\kappa {\bf E}_i \left(\sigma_3 \vec{\nabla}^2\rho+ \bar\sigma_3 \vec{\nabla}^2\rho_5 \right) -\mathcal{D}_{ij}^1 \nabla_j \rho \\
&-(\mathcal{D_{\chi}})_{ij}^1 \nabla_j \rho_5,
\end{split}
\end{equation}
\begin{equation}
\begin{split}
J^{(1)(1)}_{5i}=&\sigma_{\bar{\chi}}\kappa {\bf B}_i \delta\rho- \kappa {\bf B}_i \left(\sigma_1\vec{\nabla}^2\rho + \sigma_2\vec{\nabla}^2\rho_5 \right) -\kappa {\bf E}_i \left(\sigma_3 \vec{\nabla}^2\rho+ \bar\sigma_3 \vec{\nabla}^2\rho_5 \right) -\mathcal{D}_{ij}^1 \nabla_j \rho_5 \\
&-(\mathcal{D_{\chi}})_{ij}^1 \nabla_j \rho,
\end{split}
\end{equation}
where
\begin{align} \label{diff tensor function}
\mathcal{D}_{ij}^1&=-\delta_{ij}\kappa\left[\left(\vec{\bf B}\cdot \vec{\nabla} \right) \sigma_1 +\left(\vec{\bf E}\cdot \vec{\nabla} \right)\sigma_3 \right] +\frac{1}{4} \epsilon_{ikj} \left(\bar\rho\vec{\bf B}_k\mathcal{D}_H+2\vec{\bf E}_k\bar\sigma_{a\chi H} \right),\nonumber\\
(\mathcal{D}_\chi)_{ij}^1&=-\delta_{ij}\kappa\left[ \left(\vec{\bf B}\cdot \vec{\nabla} \right) \sigma_2+\left(\vec{\bf E}\cdot \vec{\nabla} \right)\bar\sigma_3 \right] +\frac{1}{4}\epsilon_{ikj} \left(\bar\rho_5\vec{\bf B}_k\bar{\mathcal{D}}_H +2 \vec{\bf E}_k \sigma_{a\chi H}\right).
\end{align}
$\sigma_{1,2,3}$ and $\bar\sigma_3$ constitute corrections to CME/CSE and, through spatial inhomogeneities of $\rho,\rho_5$, influence the  Ohmic conductivity.
The scalar diffusion function $\mathcal{D}$ \cite{1511.08789} now becomes tensor TCFs $\mathcal{D}_{ij}^1$ and $(\mathcal{D}_\chi)_{ij}^1$, {\it linearly} depending on $\vec{\bf E}$ and $\vec{\bf B}$ because of  the  weak field approximation adopted here.

In the hydrodynamic limit $\omega,q\ll 1$, the TCFs in (\ref{ji11}, \ref{j5i11}) are expandable (below we set $\pi T=1$ for convenience and the dimensionful frequency and momentum are $\pi T \omega$ and $\pi T q$):
\begin{equation}\label{sigmab}
\sigma_{\bar{\chi}}=6+\frac{3}{2}i\omega \left(\pi + 2\log 2\right) -\frac{1}{8} \left\{\omega^2 \left[\pi^2+6\left(4\mathcal{C}+\log^22\right)\right]+q^2 \left(12\pi-24\log 2 \right) \right\}+ \cdots,
\end{equation}
\begin{equation}
\mathcal{D}_H=\kappa^2 \left\{72(3\log2-2)+i\omega6\left[\pi(2\pi +3\log2-6) +(9\log2-12)\log2\right]+\cdots\right\},
\end{equation}
\begin{equation}
\bar{\mathcal{D}}_H=\mathcal{D}_H\left[\bar{\mu} \leftrightarrow \bar{\mu}_5\right],
\end{equation}
\begin{equation}\label{sigmaf}
\sigma_{a\chi H}= \kappa \left\{6\log2 + i\omega \frac{1}{16} \left(48\mathcal{C}+5\pi^2\right)+\cdots\right\},
\end{equation}
\begin{equation}
\bar\sigma_{a\chi H}=0+\cdots,
\end{equation}
\begin{equation}
\sigma_1=162\kappa^2\bar{\mu}\bar{\mu}_5\left[6+\log2(5\log2-12)\right]+\cdots,
\end{equation}
\begin{equation}
\sigma_2=\frac{1}{8}(6\pi-\pi^2-12\log2)+108\kappa^2(\bar{\mu}^2+\bar{\mu}_5^2) \left[6 +\log2(5\log2-12)\right]+\cdots,
\end{equation}
\begin{equation}
\sigma_3= 9\kappa\bar{\mu} \log^22+\cdots,
\end{equation}
\begin{equation}\label{sigma3b}
\bar\sigma_3=\sigma_3\left[\bar{\mu} \leftrightarrow \bar{\mu}_5\right],
\end{equation}
where $\cdots$ denotes higher powers in $\omega,q^2$ and $\mathcal{C}\approx0.915966$ is the Catalan's constant. Here, $\bar\mu=\bar\rho/2, \bar\mu_5=\bar\rho_5/2$ are backgrounds for vector/axial chemical potentials.
While each term in (\ref{sigmab}-\ref{sigma3b}) have been computed as individual TC in \cite{Bu:2018psl}, the resummation procedure here collects all relevant TCs into a single TCF and determines the most general structure of currents, valid to all orders.


Beyond the hydrodynamic limit, the TCFs are computed numerically. The results are presented and discussed in subsection \ref{s43}. We observe a relatively weak dependence on $q^2$ while $\omega$-dependence is more profound: damped oscillations towards asymptotic regime around $\omega\simeq5$. We remark that none of the TCFs survives beyond asymptotically large $\omega\gtrsim5$. For the details about each TCF we refer to subsection \ref{s43}.

It is interesting to explore dependence of the TCFs on the chemical potentials. Of special interest is the case of zero background axial charge density, $\bar\rho_5=0$, which is the most realistic scenario for any conceivable experiment.  However, even in this case, $\mu_5$ could be nonzero and would be proportional to $\vec{\bf E}\cdot\vec{\bf B}$ due to the chiral anomaly (\ref{cont eqn}). Because of the linearisation approximation,  the TCFs $\sigma_1,\bar\sigma_3$ vanish in the limit $\bar\rho_5=0$. We expect them to be nonzero beyond the current approximation.
At $q=0$, for the remaining TCFs
we discover some universal dependence:  $\bar\sigma_{a\chi H}$ vanishes; $\sigma_{a\chi H}$, $\mathcal{D}_H$, $\bar{\mathcal{D}}_H$ do not depend on the chemical potentials at all; $\sigma_1$ is linear in $\kappa^2\bar{\mu}\bar{\mu}_5$; $\sigma_3$ is linear in $\kappa\bar{\mu}$; similarly, $\bar\sigma_3$ is linear in $\kappa\bar{\mu}_5$;
$\sigma_2$ has a normal component independent of the chemical potentials and anomaly induced  correction which is linear in $\kappa^2(\bar{\mu}^2+\bar{\mu}_5^2)$.
All these features can be  derived from the underlying  equations (see  Appendix \ref{appendixb2} for relevant ODEs).

The rest of this paper is structured as follows. In Section \ref{s2}, we present the holographic model briefly. For more details about holographic model we refer to \cite{Bu:2018psl}.  Section \ref{s4} contains the main part of the study: gradient resummation for nonlinear chiral transport. It is further split into four subsections. In subsection \ref{s41},  the constitutive relations (\ref{ji11},\ref{j5i11}) are derived from the dynamical components of the bulk anomalous Maxwell equations near the conformal boundary. In subsection \ref{s42}, the TCFs are analytically computed
in the hydrodynamic limit. Subsection \ref{s43} numerically extends the results beyond this limit.
Subsection \ref{s44} focuses on the CMW dispersion relation beyond hydrodynamic limit.
Section \ref{s5} concludes our study.
Appendices supplement calculational details for Section \ref{s4}.
\section{Holographic setup: $U(1)_V\times U(1)_A$}\label{s2}
The bulk action is \cite{Yee:2009vw,Gynther:2010ed}
\begin{equation}
S=\int d^5x \sqrt{-g}\mathcal{L}+S_{\textrm{c.t.}},
\end{equation}
where
\begin{equation}\label{LPVA}
\begin{split}
\mathcal{L}=&-\frac{1}{4} (F^V)_{MN} (F^V)^{MN}-\frac{1}{4} (F^a)_{MN} (F^a)^{MN} +\frac{\kappa\,\epsilon^{MNPQR}}{2\sqrt{-g}}\\
&\times\left[3 A_M (F^V)_{NP} (F^V)_{QR} + A_M (F^a)_{NP}(F^a)_{QR}\right],
\end{split}
\end{equation}
and the counter-term action $S_{\textrm{c.t.}}$ is
\begin{equation}\label{ct VA}
S_{\textrm{c.t.}}=\frac{1}{4}\log r \int d^4x \sqrt{-\gamma}\left[(F^V)_{\mu\nu} (F^V)^{\mu\nu} +(F^a)_{\mu\nu}(F^a)^{\mu\nu}\right].
\end{equation}
The gauge Chern-Simons terms ($\sim \kappa$) in the bulk action mimic the chiral anomaly of the boundary field theory.
Note $\epsilon^{MNPQR}$ is the Levi-Civita symbol with the convention $\epsilon^{rtxyz}=+1$, while the Levi-Civita tensor is $\epsilon^{MNPQR}/\sqrt{-g}$. 

In the ingoing Eddington-Finkelstein coordinate, the metric of Schwarzschild-$AdS_5$ is
\begin{equation}
ds^2=2dtdr-r^2f(r)dt^2+r^2\delta_{ij}dx^idx^j,
\end{equation}
where $f(r)=1-1/r^4$. Here we have normalised the Hawking temperature (identified as the temperature of the boundary theory) to $\pi T=1$.

The bulk equations of motion read
\begin{equation}\label{EV}
\textrm{EV}^M\equiv \nabla_N(F^V)^{NM}+\frac{3\kappa  \epsilon^{MNPQR}} {\sqrt{-g}} (F^a)_{NP} (F^V)_{QR}=0,
\end{equation}
\begin{equation}\label{EA}
\textrm{EA}^M\equiv \nabla_N(F^a)^{NM} +\frac{3\kappa \epsilon^{MNPQR}} {2\sqrt{-g}} \left[(F^V)_{NP} (F^V)_{QR}+  (F^a)_{NP} (F^a)_{QR}\right]=0,
\end{equation}
where $\textrm{EV}^\mu=\textrm{EA}^\mu=0$ and $\textrm{EV}^r=\textrm{EA}^r=0$ correspond to dynamical and constraint equations, respectively.
The boundary currents are defined as
\begin{equation} \label{current definition}
J^\mu\equiv \lim_{r\to\infty}\frac{\delta S}{\delta V_\mu},~~~~~~~~~~~~~
J^\mu_5\equiv \lim_{r\to\infty}\frac{\delta S}{\delta A_\mu}.
\end{equation}
Employing the radial gauge $V_r=A_r=0$, it is sufficient to solve the dynamical equations only to determine the boundary currents (\ref{current definition}), leaving constraints aside. Indeed, the constraint equations give rise to continuity equations of currents (\ref{cont eqn}). Thus, without imposing the constraint equations, the currents to be constructed are {\it off-shell}.

For practical purpose, it is useful to express the currents in terms of the coefficients of near boundary ($r=\infty$)  pre-asymptotic expansion of the bulk gauge fields:
\begin{equation}\label{bdry currents}
\begin{split}
J^{\mu}	=\eta^{\mu\nu}(2V_{\nu}^{(2)}+2V^{\textrm{L}}_{\nu}+\eta^{\sigma t} \partial_{\sigma} \mathcal{F}_{t\nu}^V),\qquad \qquad
J_{5}^{\mu}= \eta^{\mu\nu}2A_{\nu}^{(2)},
\end{split}
\end{equation}
where $\mathcal{F}_{\mu\nu}^V$ is field strength of the external e/m potential $\mathcal{V}_\mu(x)$, and $4V_\mu^{\textrm{L}}=\partial^\nu \mathcal{F}_{\mu\nu}^V$. $V_\mu^{(2)}$ and $A_\mu^{(2)}$ are the coefficients of $1/r^2$ in the near boundary expansions of bulk fields $V_\mu$ and $A_\mu$, respectively. Note $V_\mu^{(2)}$ and $A_\mu^{(2)}$ have to be determined by fully solving the dynamical equations from the horizon to the boundary.

As the remainder of this section, we outline the strategy for deriving the constitutive relations for $J^\mu$ and $J_5^\mu$. To this end, we turn on finite vector/axial charge densities for the dual field theory, which are also exposed to external e/m fields $\mathcal{V}_\mu$. Holographically, the charge densities and external fields are encoded in the asymptotic behaviors of the bulk gauge fields. In the bulk, we will solve the dynamical equations assuming the charge densities and external fields as given, but without specifying them explicitly. For more details, we refer the reader to our previous publications \cite{1608.08595,1609.09054,Bu:2018psl}.

We start with the ansatz
\begin{equation} \label{corrections}
\begin{split}
V_\mu(r,x_\alpha)=\mathcal{V}_\mu(x_\alpha)-\frac{\rho(x_\alpha)}{2r^2}\delta_{\mu t}+ \mathbb{V}_\mu(r,x_\alpha),\qquad
A_\mu(r,x_\alpha)=
-\frac{\rho_{_5}(x_\alpha)}{2r^2}\delta_{\mu t} + \mathbb{A}_\mu(r,x_\alpha),
\end{split}
\end{equation}
where $\mathcal{V}_\mu(x)$ is the external gauge potential, and $\rho,\rho_5$ are vector and axial charge densities of the boundary theory.
$\mathbb{V}_\mu$ and $\mathbb{A}_\mu$ will be determined by solving dynamical equations.
Appropriate boundary conditions are classified into three types. First, $\mathbb{V}_\mu$ and $\mathbb{A}_\mu$ are regular over the domain $r\in [1,\infty)$. Second, at the conformal boundary $r=\infty$, we require
\begin{equation}\label{AdS constraint}
\mathbb{V}_\mu\to 0,~~~~~~\mathbb{A}_\mu \to 0~~~~~~~\textrm{as}~~~~~~r\to \infty,
\end{equation}
which amounts to fixing external gauge potentials to be $\mathcal{V}_\mu$ and zero (for the axial field).
Additional integration constants will be fixed by the Landau frame convention,
\begin{equation}\label{Landau frame}
J^t=\rho(x_\alpha),~~~~~~~~~~~J^t_5=\rho_{_5}(x_\alpha).
\end{equation}
The Landau frame convention corresponds to a residual gauge fixing for the bulk fields.

To facilitate the exchange between charge density and chemical potential, we define
\begin{equation} \label{def potentials}
\begin{split}
\mu&=V_t(r=\infty)-V_t(r=1)=\frac{1}{2}\rho-\mathbb{V}_t(r=1),\\
\mu_{_5}&=A_t(r=\infty)-A_t(r=1)=\frac{1}{2}\rho_{_5}-\mathbb{A}_t(r=1).
\end{split}
\end{equation}
Generically, $\mu,\mu_{_5}$ are nonlinear functionals of densities and external fields.

For generic profiles of $\mathcal{V}_\mu(x),\rho(x),\rho_5(x)$, it is impossible to solve dynamical components of (\ref{EV}, \ref{EA}). As announced in section \ref{s2}, we employ the approximation schemes (\ref{linsch1} ,\ref{linsch2}). Consequently, the corrections $\mathbb{V}_\mu$ and $\mathbb{A}_\mu$ are first expanded in powers of $\epsilon$,
\begin{equation}
\mathbb{V}_\mu= \mathbb{V}_\mu^{(0)}(r)+ \epsilon \mathbb{V}_\mu^{(1)}(r,x^\alpha)+ \mathcal{O}(\epsilon^2),\qquad\qquad
\mathbb{A}_\mu= \mathbb{A}_\mu^{(0)}(r)+ \epsilon \mathbb{A}_\mu^{(1)}(r,x^\alpha)+ \mathcal{O}(\epsilon^2),
\end{equation}
and then  each order in $\epsilon$ is further expanded in powers of $\alpha$:
\begin{equation}
\begin{split}
&\mathbb{V}_\mu^{(0)}=\sum^\infty_{n=1}\alpha^n\mathbb{V}^{(0)(n)}_\mu, \qquad \qquad  \mathbb{A}_\mu^{(0)}=\sum^\infty_{n=1}\alpha^n\mathbb{A}^{(0)(n)}_\mu, \\
&\mathbb{V}_\mu^{(1)}=\sum^\infty_{n=0}\alpha^n\mathbb{V}^{(1)(n)}_\mu, \qquad \qquad  \mathbb{A}_\mu^{(1)}=\sum^\infty_{n=0}\alpha^n\mathbb{A}^{(1)(n)}_\mu,
\end{split}
\end{equation}
where $\mathbb{V}_\mu^{(0)(1)}$, $\mathbb{V}_\mu^{(1)(0)}$, $\mathbb{A}_\mu^{(0)(1)}$, $\mathbb{A}_\mu^{(1)(0)}$ were derived in \cite{1608.08595}. Since $\mathbb{V}_\mu^{(0)(1)}$, $\mathbb{V}_\mu^{(1)(0)}$, $\mathbb{A}_\mu^{(0)(1)}$, $\mathbb{A}_\mu^{(1)(0)}$ will act as sources in the dynamical equations at $\mathcal{O}(\epsilon^1\alpha^1)$, we summarise them below (the notations here will be slightly different from \cite{1608.08595}).

At $\mathcal{O}(\epsilon^0 \alpha^1)$, we have
\begin{equation}\label{Vt01}
\mathbb{V}^{(0)(1)}_t=\mathbb{A}^{(0)(1)}_t=0,\qquad
\mathbb{V}^{(0)(1)}_i= f_1\mathbf{E}_i+f_2 \kappa \bar{\rho}_5 \mathbf{B}_i, \qquad
\mathbb{A}^{(0)(1)}_i= f_2 \kappa \bar{\rho} \mathbf{B}_i,
\end{equation}
where $f_1$ and $f_2$ are \cite{1608.08595}
\begin{equation}
f_1= -\frac{1}{4}\left[\log \frac{(1+r)^2}{1+r^2}-2\arctan(r)+\pi\right] \qquad \text{and} \qquad f_2=3 \log\frac{1+r^2}{r^2}.
\end{equation}

At $\mathcal{O}(\epsilon^1\alpha^0)$, the corrections are (note $\delta\vec E=\delta\vec B=0$ throughout this work)
\begin{equation}\label{VAti10}
\begin{split}
&\mathbb{V}^{(1)(0)}_t=g_3(r,\omega,\vec{q}\,) \delta\rho,
~~~~~~~~~~~~~~~~~~~~~~~~~~~~
\mathbb{A}^{(1)(0)}_t= g_3(r,\omega,\vec{q}\,) \delta \rho_5,\\
&\mathbb{V}^{(1)(0)}_i=g_4(r,\omega,\vec{q}\,)\partial_i \delta\rho, ~~~~~~~~~~~~~~~~~~~~~~~~~~
\mathbb{A}^{(1)(0)}_i= g_4(r,\omega,\vec{q}\,) \partial_i\delta \rho_5.
\end{split}
\end{equation}
$g_3$ and $g_4$ satisfy coupled ordinary differential equations (ODEs),
\begin{equation}
\begin{split}
&0=r^2\partial^2_r g_3+3r\partial_r g_3 -q^2\partial_r g_4,\\
&0=(r^5-r)\partial^2_r g_4+(3r^4+1)\partial_r g_4 -2i\omega r^3\partial_r g_4-i\omega r^2 g_4-r^3\partial_r g_3-r^2g_3 -\frac{1}{2},
\end{split}
\end{equation}
which were solved both analytically in the hydro limit ($\omega,q\ll1$) and numerically for generic values of $\omega,q$ in Ref. \cite{1608.08595}. 
Below we quote the hydro expansion of diffusion function $\mathcal{D}$ \cite{1511.08789} (which can be extracted from solution to $g_4$):
\begin{align}\label{D}
\mathcal{D}=\frac{1}{2}+\frac{i\omega\pi}{8}-\frac{1}{48}\left[\pi^2\omega^2-q^2
(6\log2-3\pi)\right]+\cdots,
\end{align}

At $\mathcal{O}(\epsilon^1\alpha^1)$, the dynamical equations reduce  to the following \textit{linear} partial differential equations for  the corrections $\mathbb{V}_\mu^{(1)(1)}$ and $\mathbb{A}_\mu^{(1)(1)}$:
\begin{equation}\label{Vt(11)}
\begin{split}
0&=r^3\partial_r^2\mathbb{V}^{(1)(1)}_t+3r^2\partial_r\mathbb{V}^{(1)(1)}_t+r\partial_r
\partial_k\mathbb{V}^{(1)(1)}_k +12 \kappa\epsilon^{ijk}\left(\partial_r \mathbb{A}^{(1)(0)}_i\partial_j\bar{\mathcal{V}}_k\right.\\
&\left.+\partial_r \mathbb{A}^{(0)(1)}_i\partial_j\mathbb{V}^{(1)(0)}_k +\partial_r \mathbb{V}^{(0)(1)}_i\partial_j\mathbb{A}^{(1)(0)}_k\right),
\end{split}
\end{equation}
\begin{equation}
\begin{split}
0&=(r^5-r)\partial^2_r\mathbb{V}^{(1)(1)}_i+(3r^4+1)\partial_r\mathbb{V}^{(1)(1)}_i+
2r^3\partial_r\partial_t \mathbb{V}_i^{(1)(1)}-r^3\partial_r\partial_i \mathbb{V}^{(1)(1)}_t \\ &+r^2\left(\partial_t\mathbb{V}^{(1)(1)}_i-\partial_i\mathbb{V}^{(1)(1)}_t\right)
+r\left(\partial^2\mathbb{V}^{(1)(1)}_i-\partial_i\partial_k\mathbb{V}^{(1)(1)}_k\right) +12\kappa r^2 \epsilon^{ijk}\\
&\times\left(\frac{1}{r^3}\delta \rho_5 \partial_j \bar{\mathcal{V}}_k +\frac{1}{r^3} \bar{\rho}_5\partial_j\mathbb{V}^{(1)(1)}_k+\partial_r\mathbb{A}^{(1)(0)}_t\partial_j
\bar{\mathcal{V}}_k\right) -12\kappa r^2\epsilon^{ijk}\\
&\times\left\{\partial_r \mathbb{A}^{(0)(1)}_j \left[(\partial_t\mathbb{V}^{(1)(0)}_k -\partial_k \mathbb{V}^{(1)(0)}_t)+\frac{1}{2r^2}\partial_k\delta\rho\right]+\partial_r \mathbb{A}^{(1)(0)}_j\left(\partial_t\bar{\mathcal{V}}_k-\partial_k\bar{\mathcal{V}}_t
\right)\right\}\\
&-12\kappa r^2 \epsilon^{ijk}\left\{\partial_r\mathbb{V}^{(0)(1)}_j\left[(\partial_t \mathbb{A}^{(1)(0)}_k-\partial_k\mathbb{A}^{(1)(0)}_t)+\frac{1}{2r^2}\partial_k\delta \rho_5\right]-\frac{\bar{\rho}}{r^3}\partial_j\mathbb{A}^{(1)(1)}_k\right\},
\end{split}
\end{equation}
\begin{equation}
\begin{split}
0&=r^3\partial_r^2\mathbb{A}^{(1)(1)}_t+3r^2\partial_r\mathbb{A}^{(1)(1)}_t+r\partial_r
\partial_k\mathbb{A}^{(1)(1)}_k +12 \kappa\epsilon^{ijk}\left(\partial_r \mathbb{V}^{(1)(0)}_i \partial_j\bar{\mathcal{V}}_k\right.\\
&\left.+\partial_r\mathbb{V}^{(0)(1)}_i\partial_j\mathbb{V}^{(1)(0)}_k+\partial_r
\mathbb{A}^{(0)(1)}_i\partial_j\mathbb{A}^{(1)(0)}_k\right),
\end{split}
\end{equation}
\begin{equation}\label{Ai(11)}
\begin{split}
0&=(r^5-r)\partial_r^2\mathbb{A}^{(1)(1)}_i+(3r^4+1)\partial_r\mathbb{A}^{(1)(1)}_i+2r^3
\partial_r\partial_t\mathbb{A}^{(1)(1)}_i-r^3\partial_r\partial_i\mathbb{A}^{(1)(1)}_t\\
&+r^2\left(\partial_t\mathbb{A}^{(1)(1)}_i-\partial_i\mathbb{A}^{(1)(1)}_t\right)+r\left(
\partial^2\mathbb{A}^{(1)(1)}_i-\partial_i\partial_k\mathbb{A}^{(1)(1)}_k \right) +12 \kappa r^2 \epsilon^{ijk} \\
&\times\left(\partial_j\bar{\mathcal{V}}_k(\partial_r\mathbb{V}^{(1)(0)}_t+\frac{1}{r^3}
\delta\rho)+\frac{\bar{\rho}}{r^3}\partial_j\mathbb{V}^{(1)(1)}_k\right) -12\kappa r^2 \epsilon^{ijk}\\
&\times \left\{\partial_r\mathbb{V}^{(0)(1)}_j\left[ (\partial_t \mathbb{V}^{(1)(0)}_k-\partial_k\mathbb{V}_t^{(1)(0)})+\frac{1}{2r^2}\partial_k\delta\rho
\right]+\partial_r\mathbb{V}^{(1)(0)}_j(\partial_t\bar{\mathcal{V}}_k-\partial_k\bar{
\mathcal{V}}_t)\right\} \\
&-12\kappa r^2 \epsilon^{ijk}\left\{\partial_r\mathbb{A}^{(0)(1)}_j\left[(\partial_t
\mathbb{A}^{(1)(0)}_k-\partial_k\mathbb{A}^{(1)(0)}_t)+\frac{1}{2r^2}\partial_k\delta
\rho_5\right]-\frac{\bar{\rho_5}}{r^3}\partial_j\mathbb{A}^{(1)(1)}_k\right\}.
\end{split}
\end{equation}
In the next section \ref{s4}, we will solve (\ref{Vt(11)}-\ref{Ai(11)}) by the technique invented in \cite{1406.7222,1409.3095}.

\section{Nonlinear chiral transport and gradient resummation}\label{s4}

In this section, we focus on all order gradient resummation. It is split into four subsections. The first one \ref{s41} is devoted to derivation of the constitutive relations (\ref{ji11}, \ref{j5i11}). In the following subsections \ref{s42} and \ref{s43},
 the TCFs in (\ref{ji11}, \ref{j5i11}) are evaluated, first
analytically in the hydrodynamic limit, and then numerically for arbitrary momenta. The last subsection \ref{s44} is about non-dissipative modes in the CMW dispersion relations.

\subsection{Constitutive relations at  $\mathcal{O}(\epsilon^1\alpha^1)$}\label{s41}

Following the formalism introduced in  \cite{1406.7222,1409.3095}, the corrections $\mathbb{V}_\mu^{(1)(1)}$ and $\mathbb{A}_\mu^{(1)(1)}$ are decomposed in terms of basic structures built from the external fields and inhomogeneous parts of the charge densities,
\begin{equation}\label{decomposition Vt11}
\mathbb{V}^{(1)(1)}_t=S_1 \kappa \mathbf{B}_k\partial_k \delta \rho +S_2 \kappa \mathbf{E}_k \partial_k \delta \rho +S_3 \kappa \mathbf{B}_k\partial_k \delta \rho_5 +S_4 \kappa \mathbf{E}_k \partial_k \delta \rho_5,
\end{equation}
\begin{equation} \label{decomposition Vi11}
\begin{split}
\mathbb{V}^{(1)(1)}_i&=V_1\kappa \mathbf{B}_i \delta \rho +V_2\kappa \mathbf{B}_k \partial_i \partial_k \delta \rho +V_3\kappa \epsilon^{ijk} \mathbf{B}_j \partial_k \delta \rho +V_4\kappa \mathbf{E}_i \delta \rho +V_5\kappa \mathbf{E}_k \partial_i \partial_k \delta \rho , \\
&+V_6\kappa \epsilon^{ijk} \mathbf{E}_j \partial_k \delta \rho+V_7\kappa \mathbf{B}_i \delta \rho_5 +V_8\kappa \mathbf{B}_k \partial_i \partial_k \delta \rho_5 +V_9\kappa \epsilon^{ijk} \mathbf{B}_j \partial_k \delta \rho_5 +V_{10} \kappa \mathbf{E}_i \delta \rho_5 \\
&+V_{11}\kappa \mathbf{E}_k \partial_i \partial_k \delta \rho_5 +V_{12}\kappa \epsilon^{ijk} \mathbf{E}_j \partial_k \delta \rho_5,
\end{split}
\end{equation}
\begin{equation} \label{decomposition At11}
\mathbb{A}^{(1)(1)}_t=\bar{S}_1 \kappa\mathbf{B}_k\partial_k \delta \rho +\bar{S}_2 \kappa \mathbf{E}_k \partial_k \delta \rho +\bar{S}_3\kappa \mathbf{B}_k\partial_k \delta \rho_5 +\bar{S}_4\kappa \mathbf{E}_k \partial_k \delta \rho_5,
\end{equation}
\begin{equation} \label{decomposition Ai11}
\begin{split}
\mathbb{A}^{(1)(1)}_i&=\bar{V}_1 \kappa\mathbf{B}_i \delta \rho +\bar{V}_2\kappa \mathbf{B}_k \partial_i \partial_k \delta \rho +\bar{V}_3\kappa \epsilon^{ijk} \mathbf{B}_j \partial_k \delta \rho +\bar{V}_4\kappa \mathbf{E}_i \delta \rho +\bar{V}_5 \kappa \mathbf{E}_k \partial_i \partial_k \delta \rho  \\
&+\bar{V}_6\kappa \epsilon^{ijk} \mathbf{E}_j \partial_k \delta \rho+\bar{V}_7\kappa \mathbf{B}_i \delta \rho_5 +\bar{V}_8 \kappa\mathbf{B}_k \partial_i \partial_k \delta \rho_5 +\bar{V}_9\kappa \epsilon^{ijk} \mathbf{B}_j \partial_k \delta \rho_5 +\bar{V}_{10}\kappa \mathbf{E}_i \delta \rho_5 \\
&+\bar{V}_{11}\kappa \mathbf{E}_k \partial_i \partial_k \delta \rho_5 +\bar{V}_{12} \kappa \epsilon^{ijk} \mathbf{E}_j \partial_k \delta \rho_5,
\end{split}
\end{equation}
where $S_i$, $\bar{S}_i$, $V_i$ and $\bar{V}_i$ are functionals of the boundary derivative operator $\partial_\mu$ and functions of the radial coordinate $r$.
They also depend on the constant values $\bar\mu$ and $\bar\mu_5$ of the chemical potentials.  Fourier transforming $\delta\rho $ and $\delta\rho_5$  turns all the derivatives
into momenta. Thus, in momentum space, these decomposition coefficients become functions of the radial coordinate, frequency $\omega$ and spatial momentum squared $q^2$:
\begin{eqnarray}
&&S_i\left(r,\partial_t,\partial_i^2\right)\rightarrow S_i\left(r,\omega,q^2\right) \qquad \qquad \bar{S}_i\left(r,\partial_t,\partial_i^2\right)\rightarrow \bar{S}_i\left(r,\omega,q^2\right),\\
&&V_i\left(r,\partial_t,\partial_i^2\right)\rightarrow V_i\left(r,\omega,q^2\right) \qquad \qquad \bar{V}_i\left(r,\partial_t,\partial_i^2\right)\rightarrow \bar{V}_i\left(r,\omega,q^2\right),
\end{eqnarray}
which satisfy partially decoupled inhomogeneous ODEs listed in Appendix-\ref{appendixb2}.  The decomposition functions $S_i$, $\bar{S}_i$, $V_i$ and $\bar{V}_i$  are nothing
else but elements of the inverse Green function matrix for the system of ODEs.

As  discussed in Section \ref{s2}, the boundary conditions for the decomposition coefficients in (\ref{decomposition Vt11}-\ref{decomposition Ai11}) are
\begin{eqnarray}
S_i\rightarrow 0, \qquad\quad \bar{S}_i\rightarrow 0 \qquad\quad V_i\rightarrow 0, \qquad\quad \bar{V}_i\rightarrow 0 \qquad \qquad \text{as} \qquad r\rightarrow\infty. \label{bc1} \\
S_i, \quad \bar{S}_i,\quad V_i,\quad \bar{V}_i \qquad \text{are regular over the whole integral of} ~ r\in[1,\infty).\label{bc2}
\end{eqnarray}
Additional integration constants will be fixed by the Landau  frame convention (\ref{Landau frame}).

Solving the ODEs (\ref{ODEb}-\ref{ODEf}) near the boundary $r=\infty$ reveals the pre-asymptotic behaviour for the corrections, which can be summarised as
\begin{equation}\label{preasymp}
\begin{split}
S_i\rightarrow\frac{s_{i}^{1}}{r}+\frac{s_i}{r^2}+\frac{s_{i}^{\rm L}\log r}{r^2} +\cdots, \qquad  \quad \quad V_i\rightarrow\frac{v_{i}^{1}}{r}+\frac{v_i}{r^2}+\frac{v_{i}^{\rm L}\log r}{r^2} +\cdots, \\
\bar{S}_i\rightarrow\frac{\bar{s}_{i}^{1}}{r}+\frac{\bar{s}_i}{r^2}+\frac{\bar{s}_{i}^{\rm L}\log r}{r^2}+\cdots, \qquad  \quad \quad \bar{V}_i\rightarrow\frac{\bar{v}_{i}^{1}}{r}+\frac{\bar{v}_i}{r^2}+\frac{\bar{v}_{i}^{\rm L}\log r}{r^2}+\cdots,
\end{split}
\end{equation}
where $s_{i}^{1,{\rm L}}$, $v_{i}^{1,{\rm L}}$, $\bar{s}_{i}^{1,{\rm L}}$, $\bar{v}_{i}^{1,{\rm L}}$ are  fixed uniquely from the near-boundary analysis alone, while the coefficients $s_i$, $v_i$, $\bar{s}_i$, $\bar{v}_i$ can be determined only when the ODEs are fully solved in the entire bulk, from the horizon to the $AdS$ boundary.

Then, at $\mathcal{O}(\epsilon^1\alpha^1)$ the boundary currents  (\ref{bdry currents}) are
\begin{equation} \label{jt11 pre}
J^{t(1)(1)}=-2\kappa\left(s_1 \mathbf{B}_k\partial_k \delta \rho +s_2 \mathbf{E}_k \partial_k \delta \rho +s_3 \mathbf{B}_k\partial_k \delta \rho_5 +s_4 \mathbf{E}_k \partial_k \delta \rho_5\right),
\end{equation}
\begin{equation} \label{ji11 pre}
\begin{split}
J^{i(1)(1)}&=2\kappa\left(v_1 \mathbf{B}_i \delta \rho +v_2 \mathbf{B}_k \partial_i \partial_k \delta \rho +v_3 \epsilon^{ijk} \mathbf{B}_j \partial_k \delta \rho +v_4 \mathbf{E}_i \delta \rho +v_5 \mathbf{E}_k \partial_i \partial_k \delta \rho \right. \\
&\left. +v_6 \epsilon^{ijk} \mathbf{E}_j \partial_k \delta \rho+v_7 \mathbf{B}_i \delta \rho_5 +v_8 \mathbf{B}_k \partial_i \partial_k \delta \rho_5 +v_9 \epsilon^{ijk} \mathbf{B}_j \partial_k \delta \rho_5 +v_{10} \mathbf{E}_i \delta \rho_5\right.\\
&\left. +v_{11} \mathbf{E}_k \partial_i \partial_k \delta \rho_5 +v_{12} \epsilon^{ijk} \mathbf{E}_j \partial_k \delta \rho_5\right),
\end{split}
\end{equation}
\begin{equation} \label{j5t11 pre}
J_5^{t(1)(1)}=-2\kappa\left(\bar{s}_1 \mathbf{B}_k\partial_k \delta \rho +\bar{s}_2 \mathbf{E}_k \partial_k \delta \rho +\bar{s}_3 \mathbf{B}_k\partial_k \delta \rho_5 +\bar{s}_4 \mathbf{E}_k \partial_k \delta \rho_5\right),
\end{equation}
\begin{equation} \label{j5i11 pre}
\begin{split}
J_5^{i(1)(1)}&=2\kappa\left(\bar{v}_1 \mathbf{B}_i \delta \rho +\bar{v}_2 \mathbf{B}_k \partial_i \partial_k \delta \rho +\bar{v}_3 \epsilon^{ijk} \mathbf{B}_j \partial_k \delta \rho +\bar{v}_4 \mathbf{E}_i \delta \rho +\bar{v}_5 \mathbf{E}_k \partial_i \partial_k \delta \rho \right.  \\
&\left.+\bar{v}_6 \epsilon^{ijk} \mathbf{E}_j \partial_k \delta \rho+\bar{v}_7 \mathbf{B}_i \delta \rho_5 +\bar{v}_8 \mathbf{B}_k \partial_i \partial_k \delta \rho_5 +\bar{v}_9 \epsilon^{ijk} \mathbf{B}_j \partial_k \delta \rho_5 +\bar{v}_{10} \mathbf{E}_i \delta \rho_5 \right.\\
&\left.+\bar{v}_{11} \mathbf{E}_k \partial_i \partial_k \delta \rho_5 +\bar{v}_{12} \epsilon^{ijk} \mathbf{E}_j \partial_k \delta \rho_5\right).
\end{split}
\end{equation}
The Landau frame convention (\ref{Landau frame}) implies
\begin{equation}\label{si=0}
s_i=\bar{s}_i=0,~~~~~~~i=1,2,3,4.
\end{equation}

Combined with the ODEs (\ref{ODEb}, \ref{ODEf}), (\ref{si=0}) leads to constraints among the decomposition coefficients in (\ref{decomposition Vt11}-\ref{decomposition Ai11}), see (\ref{ident1}, \ref{ident2}, \ref{constraints}). Helped by these constraints, (\ref{ji11 pre}, \ref{j5i11 pre}) can be eventually put into  compact form (\ref{ji11}, \ref{j5i11}).
All the TCFs can be identified with  the near boundary data $v_i,\bar{v}_i$:
\begin{align} \label{bdry data to transports}
& \sigma_{\bar{\chi}}= 2\left(\bar{v}_1-q^2\bar{v}_2\right), \qquad  -\frac{\bar\rho}{4\kappa} \mathcal{D}_H = 2v_3 =2\bar{v}_9, \qquad -\frac{\bar\rho_5}{4\kappa} \bar{\mathcal{D}}_H =2v_9=2\bar{v}_3, \nonumber\\
&-\frac{1}{2\kappa}\sigma_{a\chi H}=2v_{12}=2\bar{v}_6,\qquad -\frac{1}{2\kappa}\bar\sigma_{a\chi H} =2v_6=2\bar{v}_{12},\qquad \sigma_1=2v_2=2\bar{v}_8, \nonumber \\ &\sigma_2=2v_8=2\bar{v}_2,\qquad \sigma_3=2v_5=2\bar{v}_{11},\qquad  \bar\sigma_3=2v_{11}=2\bar{v}_5.
\end{align}
The TCF $\sigma_{\bar{\chi}}$ does not depend on $\bar{\mu}, \bar{\mu}_5$ at all.
The rest of the TCFs bear reminiscence of the axial symmetry. It get reflected in some mirror symmetries
with respect to exchange of $\bar\rho$ and ${\bar\rho}_5$ (or equivalently of $\bar{\mu} \leftrightarrow\bar{\mu}_5$).
We found some ``symmetric relations" among the decomposition coefficients in (\ref{decomposition Vt11}-\ref{decomposition Ai11}), see (\ref{symmetric1}, \ref{symmetric2}). Consequently, the TCFs satisfy
\begin{align} \label{symmetric transports}
&\sigma_{1,2}\left[\bar{\mu},\bar{\mu}_5\right]=\sigma_{1,2}\left[\bar{\mu}_5, \bar{\mu} \right],\qquad \sigma_{a\chi H}\left[\bar{\mu},\bar{\mu}_5\right]= \sigma_{a\chi H}\left[\bar{\mu}_5, \bar{\mu} \right],\qquad \bar{\sigma}_{a\chi H} \left[\bar{\mu},\bar{\mu}_5\right]= \bar{\sigma}_{a\chi H}\left[\bar{\mu}_5, \bar{\mu} \right],\nonumber\\
&\bar{\mathcal{D}}_H[\bar\mu,\bar\mu_5]= \mathcal{D}_H[\bar\mu,\bar\mu_5]|_{\bar{\mu} \leftrightarrow \bar{\mu}_5},\qquad
\bar\sigma_3[\bar\mu,\bar\mu_5]= \sigma_3[\bar\mu,\bar\mu_5] |_{\bar{\mu}\leftrightarrow \bar{\mu}_5 }.
\end{align}

Instead of the charge densities $\rho,\rho_5$, chemical potentials are frequently used as hydrodynamic variables to parameterise the currents' constitutive relations. Up to $\mathcal{O}(\epsilon^1\alpha^1)$, the chemical potentials defined in (\ref{def potentials}) are
\begin{equation}\label{et}
\begin{split}
&\mu =\frac{1}{2}\rho(x_{\alpha})-\left[g_3(r=1) \delta \rho+\kappa\bar{S}_1(r=1) \mathbf{B}_k \partial_k\delta\rho_5\right], \\
&\mu_5=\frac{1}{2}\rho_5(x_{\alpha})-\left[g_3(r=1) \delta \rho_5+ \kappa \bar{S}_1(r=1) \mathbf{B}_k \partial_k\delta\rho\right],
\end{split}
\end{equation}
where $g_3(r=1)$ and $\bar{S}_1(r=1)$ denote horizon values of $g_3$ (appearing in (\ref{VAti10})) and $\bar{S}_1$, respectively. Then, (\ref{et}) can be inverted
\begin{equation}
\rho=\frac{1}{\frac{1}{2}-g_3(r=1)} \mu+ \frac{\kappa \bar{S}_1(r=1) B_k \partial_k } {\left[ \frac{1}{2}- g_3(r=1)\right]^2} \mu_5,\qquad \qquad
\rho_5=\frac{1}{\frac{1}{2}-g_3(r=1)} \mu_5+ \frac{\kappa \bar{S}_1(r=1) B_k \partial_k }{\left[ \frac{1}{2}- g_3(r=1)\right]^2} \mu,
\end{equation}
where we have utilised the fact that $g_3$ has non-vanishing value starting from second order in the gradient counting. After some manipulations, the currents (\ref{ji11}, \ref{j5i11}) turn into
\begin{eqnarray}\label{ji11 mu}
\vec{J}^{\;(1)(1)}=&& \sigma_{\bar{\chi}}^\prime \kappa \vec{\bf{B}} \delta\mu_5 -\frac{1}{4} \mathcal{D}_H^\prime (\bar\rho\vec{\bf{B}}\times\vec{\nabla}\delta\mu) - \frac{1}{4} \bar{\mathcal{D}}_H^\prime (\bar\rho_5\vec{\bf{B}}\times\vec{\nabla}\delta \mu_5)-\frac{1}{2}\sigma_{a\chi H}^\prime (\vec{\bf{E}} \times\vec{\nabla} \delta\mu_5) \nonumber\\
&&-\frac{1}{2} \bar{\sigma}_{a\chi H}^\prime (\vec{\bf{E}}\times\vec{\nabla}\delta\mu)
+\sigma_1^\prime \kappa \left[(\vec{\bf{B}}\times\vec{\nabla})\times\vec{\nabla}\right] \delta\mu++ \sigma_2^\prime \kappa \left[(\vec{\bf{B}}\times\vec{\nabla})\times\vec{\nabla}\right] \delta\mu_5\nonumber\\
&&+ \sigma_3^\prime \kappa \left[(\vec{\bf{E}}\times\vec{\nabla})\times\vec{\nabla} \right] \delta\mu + \bar\sigma_3^\prime \kappa\left[(\vec{\bf{E}}\times\vec{\nabla})\times\vec{\nabla}\right] \delta\mu_5,
\end{eqnarray}
\begin{eqnarray}\label{j5i11 mu5}
\vec{J}_5^{\;(1)(1)}=&&\sigma_{\bar{\chi}}^\prime \kappa \vec{\bf{B}} \delta\mu- \frac{1}{4}\mathcal{D}_H^\prime (\bar\rho\vec{\bf{B}}\times\vec{\nabla}\delta\mu_5) - \frac{1}{4}\bar{\mathcal{D}}_H^\prime (\bar\rho_5\vec{\bf{B}}\times\vec{\nabla} \delta\mu)-\frac{1}{2} \sigma_{a\chi H}^\prime (\vec{\bf{E}}\times\vec{\nabla}\delta\mu) \nonumber \\
&&- \frac{1}{2} \bar{\sigma}_{a\chi H}^\prime(\vec{\bf{E}}\times\vec{\nabla}\delta\mu_5) +\sigma_1^\prime \kappa \left[(\vec{\bf{B}}\times\vec{\nabla})\times\vec{\nabla}\right] \delta\mu_5+ \sigma_2^\prime \kappa \left[(\vec{\bf{B}}\times\vec{\nabla})\times\vec{\nabla}\right]\delta\mu_5 \nonumber\\
&&+ \sigma_3^\prime \kappa \left[(\vec{\bf{E}}\times\vec{\nabla})\times\vec{\nabla}\right] \delta\mu+ \bar\sigma_3^\prime \kappa\left[(\vec{\bf{E}}\times\vec{\nabla})\times\vec{\nabla}\right] \delta\mu ,
\end{eqnarray}
where the TCFs with prime are related to those in (\ref{ji11}, \ref{j5i11}) by
\begin{equation}\label{mutorho}
\sigma^\prime_{\bar{\chi}}= \frac{\sigma_{\bar{\chi}}}{\frac{1}{2}-g_3(r=1)},
\end{equation}
and similar equations for the rest.

\subsection{Hydrodynamic expansion: analytical results}\label{s42}

In the hydrodynamic limit $\omega,q\ll1$, the ODEs (\ref{ODEb}-\ref{ODEf}) can be solved  perturbatively. We employ the expansion parameter $\lambda$ via $\left(\omega,\vec{q}\right) \to \left(\lambda \omega, \lambda \vec{q}\right)$. Then, the decomposition coefficients are expanded in powers of $\lambda$,
\begin{equation}\label{pertdec}
\begin{split}
S_i=\sum^{\infty}_{n=0} \lambda^n S_i^{(n)}, \qquad \qquad \qquad \qquad V_i=\sum^{\infty}_{n=0} \lambda^n V_i^{(n)},\\
\bar{S}_i=\sum^{\infty}_{n=0} \lambda^n \bar{S}_i^{(n)}, \qquad \qquad \qquad \qquad \bar{V}_i=\sum^{\infty}_{n=0} \lambda^n \bar{V}_i^{(n)}.
\end{split}
\end{equation}
Then, at each order in $\lambda$, the solutions are expressed as double integrals over $r$, see Appendix-\ref{pertubativs}. The hydrodynamic expansions of $v_i$ and $\bar{v}_i$ (\ref{preasymp}) can be directly read off from (\ref{pertb}-\ref{pertf}).
Plugging these results into (\ref{bdry data to transports}) leads to the hydrodynamic expansion of all the TCFs in (\ref{ji11}, \ref{j5i11}), as presented in (\ref{sigmab}-\ref{sigma3b}).

\subsection{Beyond the hydrodynamic limit: numerical results}\label{s43}

In this section, we present our results for the TCFs in (\ref{ji11}, \ref{j5i11}) for finite frequency/momentum via solving the ODEs (\ref{ODEb}-\ref{ODEf}) numerically. Pseudo-spectral collation method is employed, which essentially converts the continuous boundary value problem of linear ODEs into that of discrete linear algebra.
For more details on the numerical method, we recommend the references \cite{spectral1,spectral2,spectral3}.
Thanks to the symmetry relations (\ref{symmetric transports}), we plot the TCFs $\sigma_{a\chi H}$, $\bar\sigma_{a\chi H}$, $\sigma_{1,2}$ for $\kappa \bar{\mu} \geqslant \kappa \bar{\mu}_5$  only without loss of generality. For $\mathcal{D}_H$ and $\sigma_3$,  this constraint is abandoned so that $\bar{\mathcal{D}}_H$ and $\bar\sigma_3$ could be extracted from $\mathcal{D}_H$ and $\sigma_3$
via the exchange $\bar{\mu}\leftrightarrow \bar{\mu}_5$.

\begin{figure}[htbp]
\centering
\includegraphics[width=\textwidth]{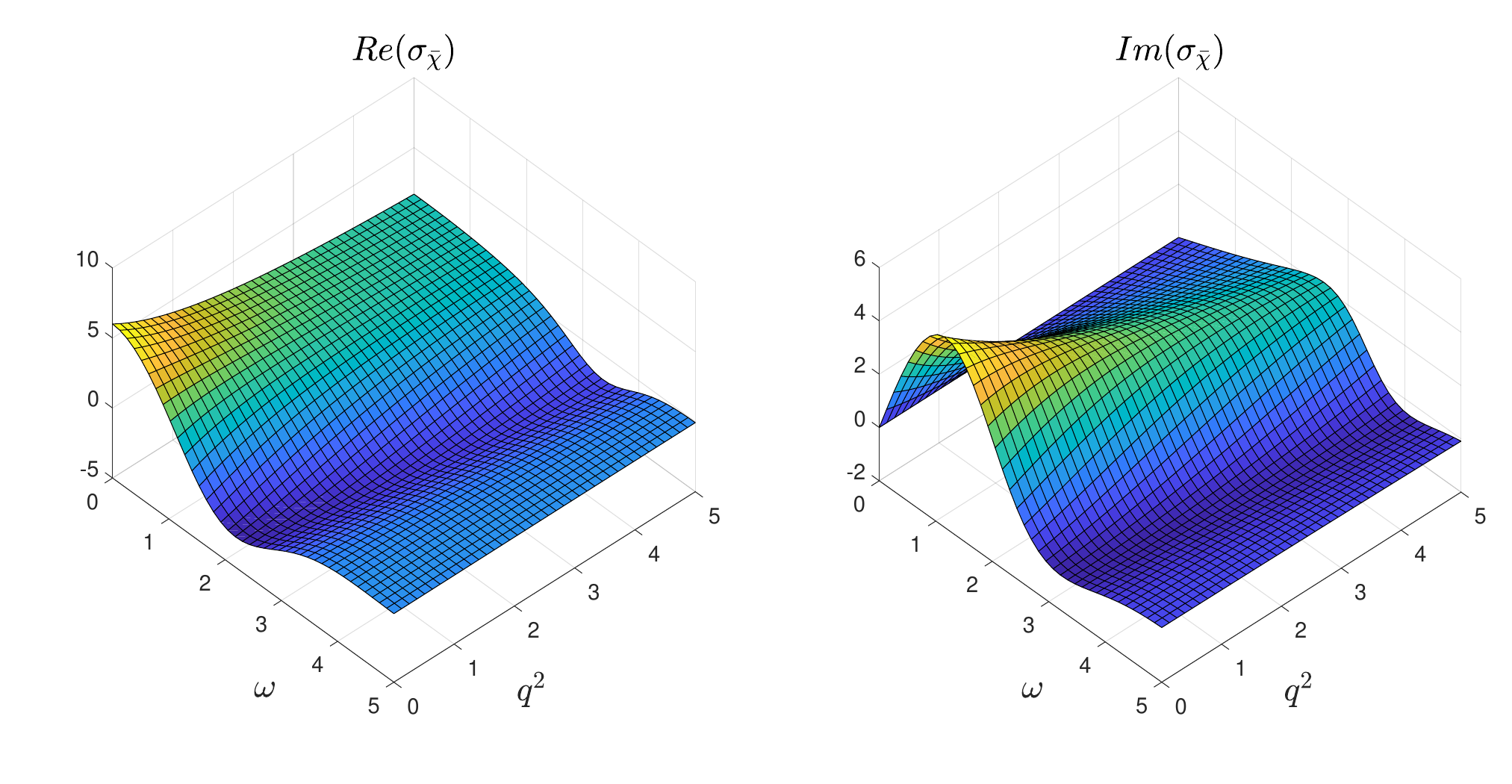}
\caption{The generalised CME/CSE conductivity $\sigma_{\bar{\chi}}$ as a function of $\omega$ and $q^2$.}\label{sigma_chixa}
\end{figure}

\begin{figure}[htbp]
\centering
\includegraphics[width=\textwidth]{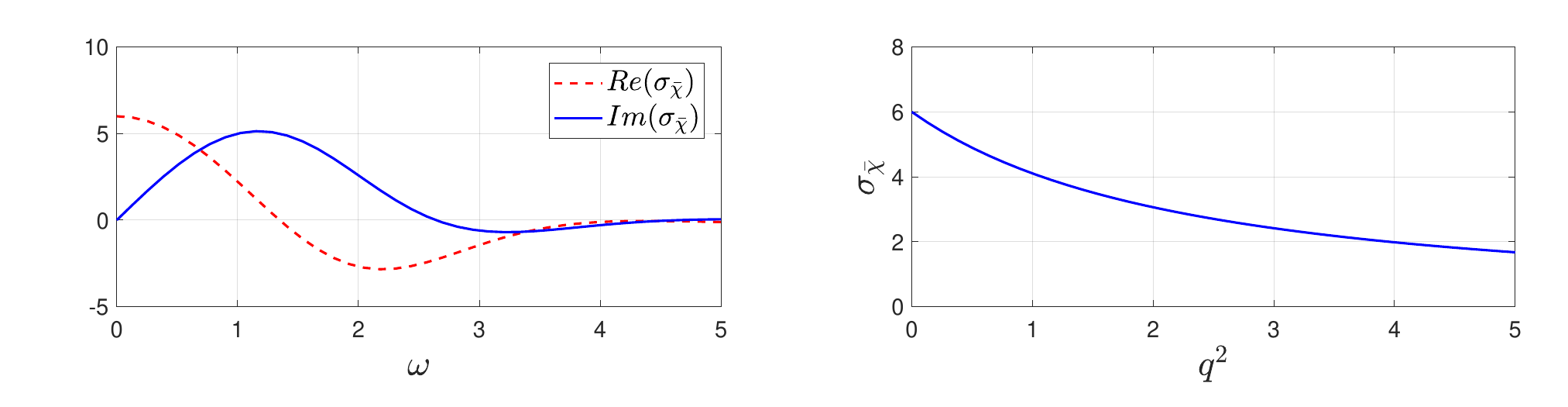}
\caption{$\omega$-dependence of $\sigma_{\bar{\chi}}$ when $q=0$ (left); $q^2$-dependence of $\sigma_{\bar{\chi}}$ when $\omega=0$ (right).}\label{sigma_chixc}
\end{figure}

First, consider TCF $\sigma_{\bar{\chi}}$, which  generalises the original CME (CSE) and measures the response to inhomogeneity of charge density $\rho$ ($\rho_5$). Note $\sigma_{\bar{\chi}}$ does not depend on the vector/axial chemical potentials at all, as can be seen from the relevant  ODEs (\ref{eom Sb1},\ref{eom Vb1},\ref{eom Vb2}). In Figure \ref{sigma_chixa} we show the 3D plot of $\sigma_{\bar{\chi}}$. The plots in Figure \ref{sigma_chixc} are 2D slices of Figure \ref{sigma_chixa} when either $\omega=0$ or $q=0$. While $\sigma_{\bar{\chi}}$ is different from the chiral magnetic conductivity $\sigma_\chi$ of \cite{1608.08595}, it has roughly the same dependence on frequency/momentum as $\sigma_\chi$ as is clear from these plots. Namely, $\sigma_{\bar{\chi}}$ shows a relatively weak dependence on $q^2$ while its dependence on $\omega$ is more profound: damped oscillations towards asymptotic regime around $\omega\simeq5$ where $\sigma_{\bar{\chi}}$ vanishes essentially. As will be clear later, this damped oscillating behavior is also observed in all other TCFs. This phenomenon can be related to quasi-normal modes in the presence of background fields, but here we are not pursuing this connection any further. When $q=0$ we computed the inverse Fourier transform of $\sigma_{\bar{\chi}}$, that is the memory function $\tilde{\sigma}_{\bar{\chi}}(t)$ of (\ref{sigmab_t}), as displayed in Figure \ref{memory}.


Next we consider TCFs $\mathcal{D}_H$, $\bar{\mathcal{D}}_H$, $\sigma_{a\chi H}$ and $\bar{\sigma}_{a\chi H}$ multiplying second order derivative structures. These second order derivative structures are cross products between electric/magnetic fields and gradient of the densities. Via the crossing rule (\ref{symmetric transports}), the Hall diffusion functions $\mathcal{D}_H$ and $\bar{\mathcal{D}}_H$ satisfy $\bar{\mathcal{D}}_H=\mathcal{D}_H(\bar{\mu} \leftrightarrow \bar{\mu}_5)$. Thus, we will mainly focus on $\mathcal{D}_H$. $\sigma_{a\chi H}$ is the anomalous chiral Hall TCF and $\bar{\sigma}_{a\chi H}$ is its axial analogue. Since $V_4=q^2V_5$ and $\bar{V}_4=q^2 \bar{V}_5$ (see (\ref{constraints})), from the ODE (\ref{eom V6}) it is obvious that $\bar{\sigma}_{a\chi H}$ has an overall $q^2$ factor, so we will plot $\bar{\sigma}_{a\chi H}/q^2$ in order to see non-trivial behavior.

\begin{figure}[htbp]
\centering
\includegraphics[width=\textwidth]{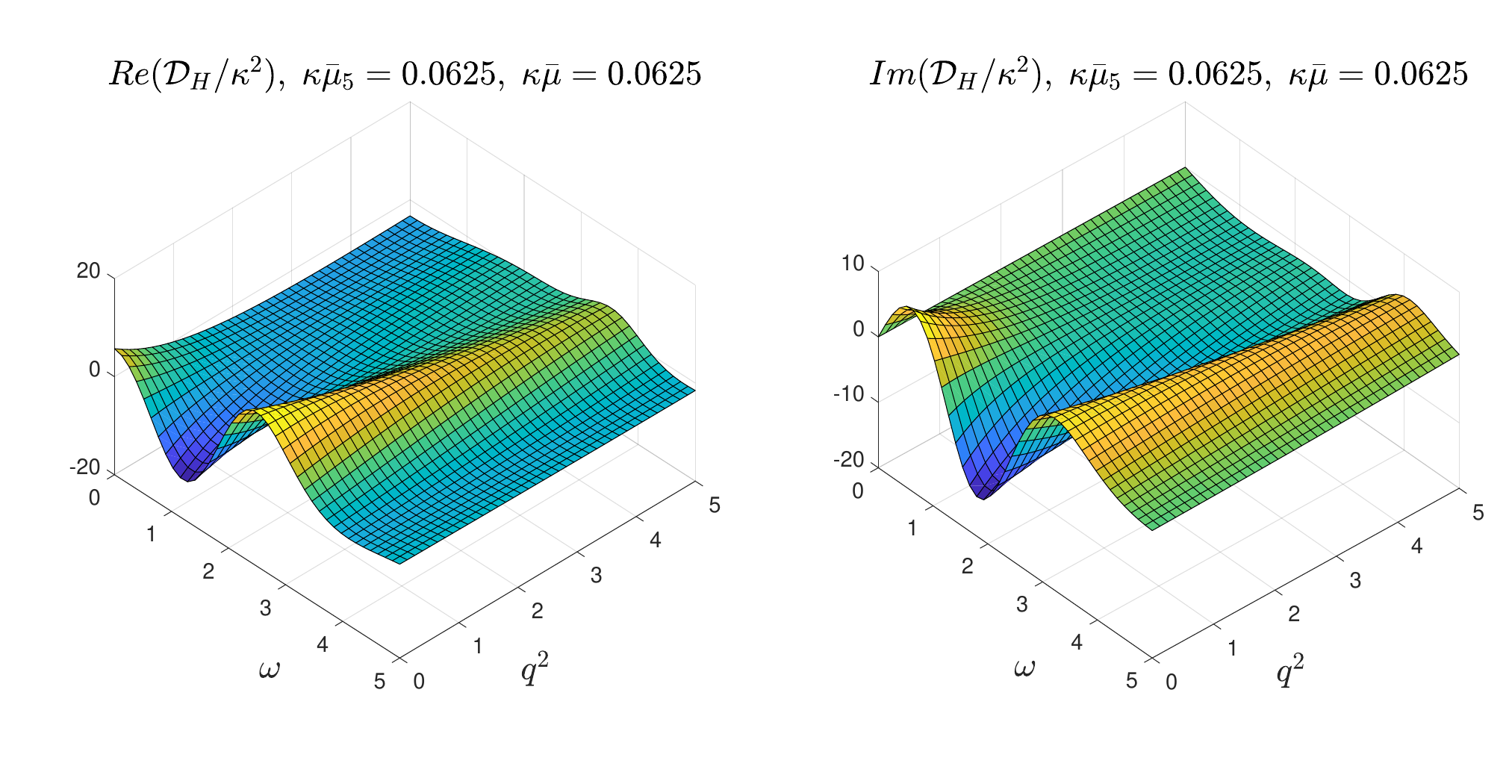}
\caption{Hall diffusion TCF $\mathcal{D}_H/\kappa^2$ as a function of $\omega$ and $q^2$ when $\kappa\bar{\mu}=\kappa\bar{\mu}_5=1/16$.}\label{sigma_2xa}
\end{figure}

\begin{figure}[htbp]
\centering
\includegraphics[width=\textwidth]{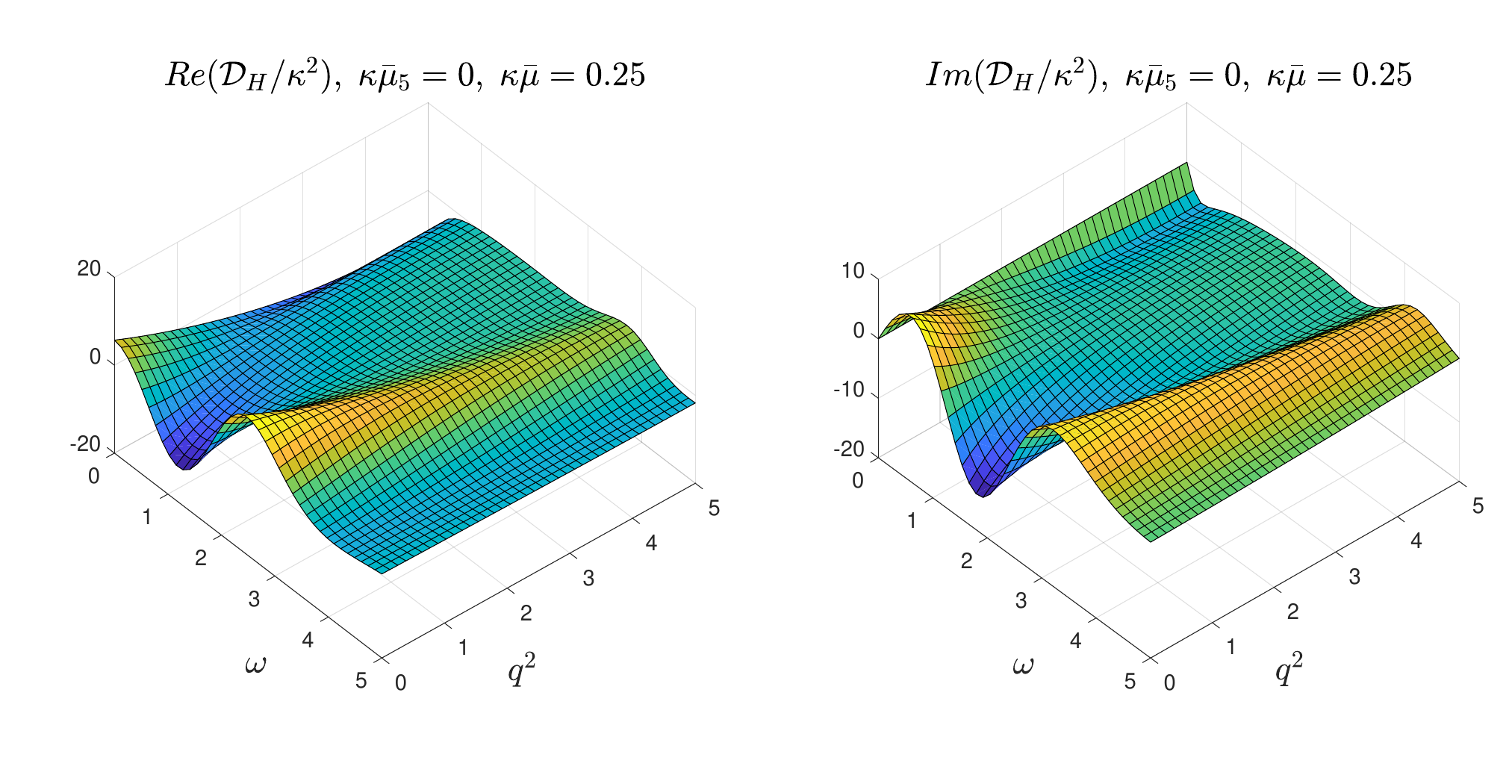}
\caption{Hall diffusion TCF $\mathcal{D}_H/\kappa^2$ as a function of $\omega$ and $q^2$ when $\kappa\bar{\mu}=1/4$, $\kappa\bar{\mu}_5=0$.}\label{sigma_2xb}
\end{figure}

\begin{figure}[htbp]
\centering
\includegraphics[width=\textwidth]{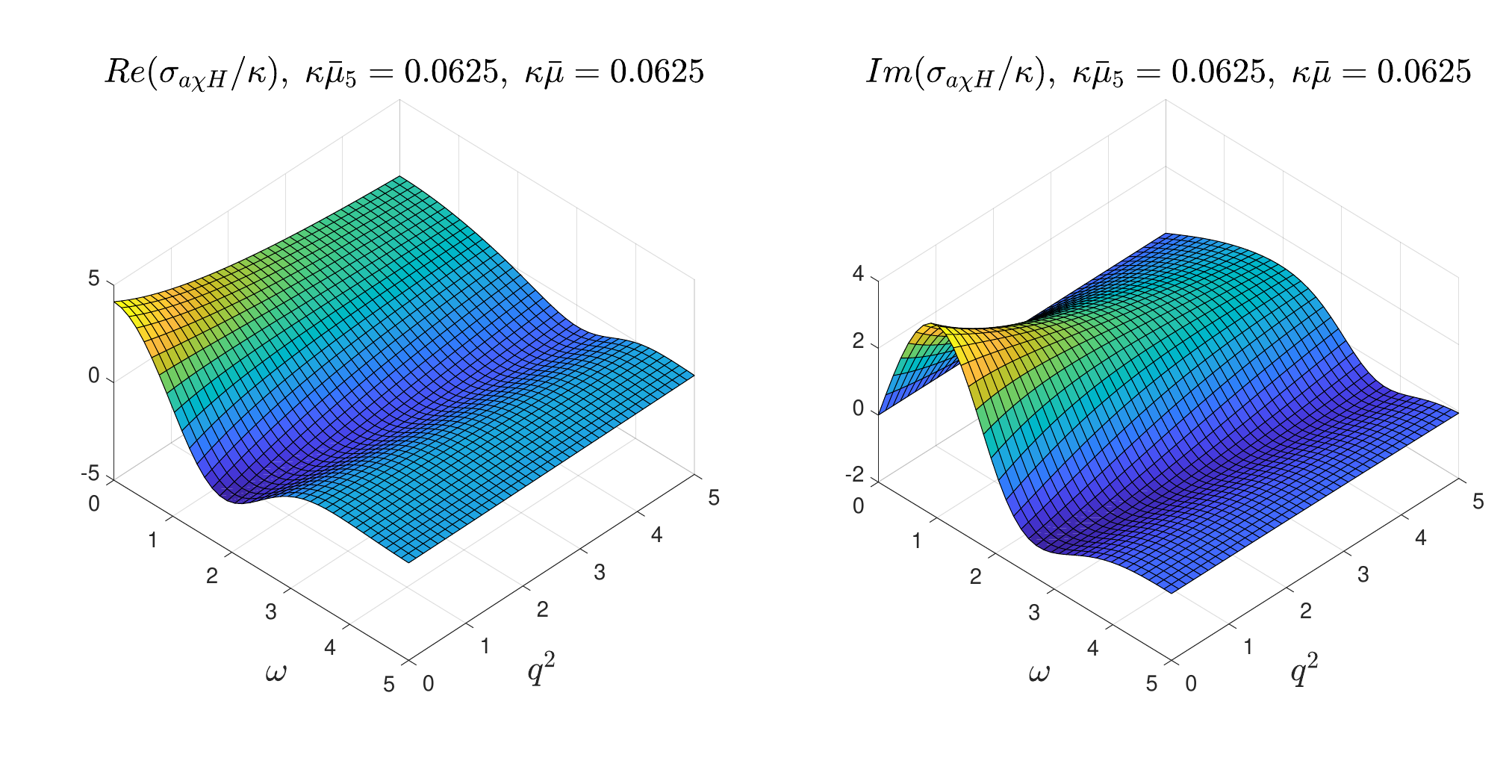}
\caption{Anomalous chiral Hall TCF $\sigma_{a\chi H}/\kappa$ as a function of $\omega$ and $q^2$ when $\kappa\bar{\mu}=\kappa\bar{\mu}_5=1/16$.}\label{sigma_8xa}
\end{figure}

\begin{figure}[htbp]
\centering
\includegraphics[width=\textwidth]{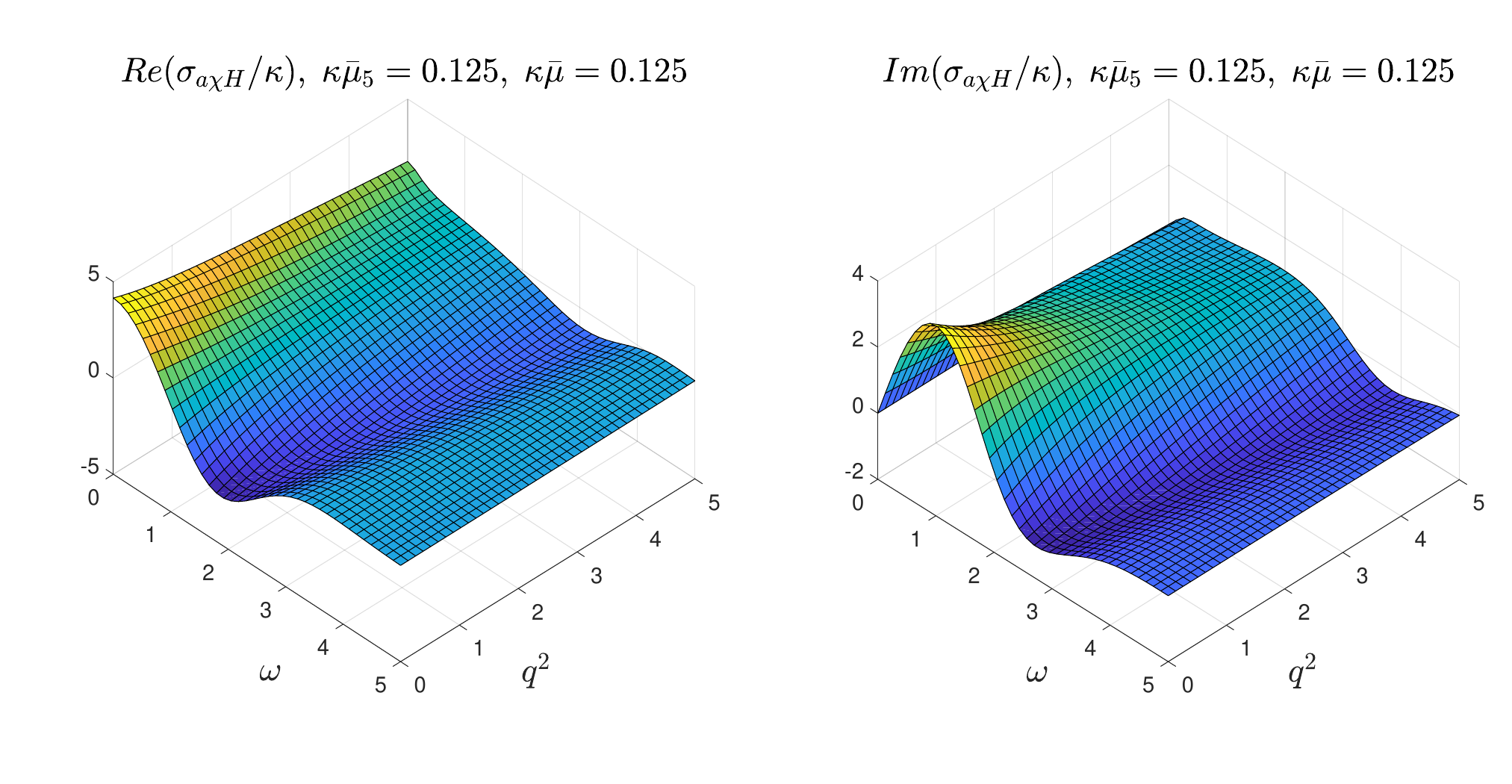}
\caption{Anomalous chiral Hall TCF $\sigma_{a\chi H}/\kappa$ as a function of $\omega$ and $q^2$ when $\kappa\bar{\mu}=\kappa\bar{\mu}_5=1/8$.}\label{sigma_8xa2}
\end{figure}

\begin{figure}[htbp]
\centering
\includegraphics[width=\textwidth]{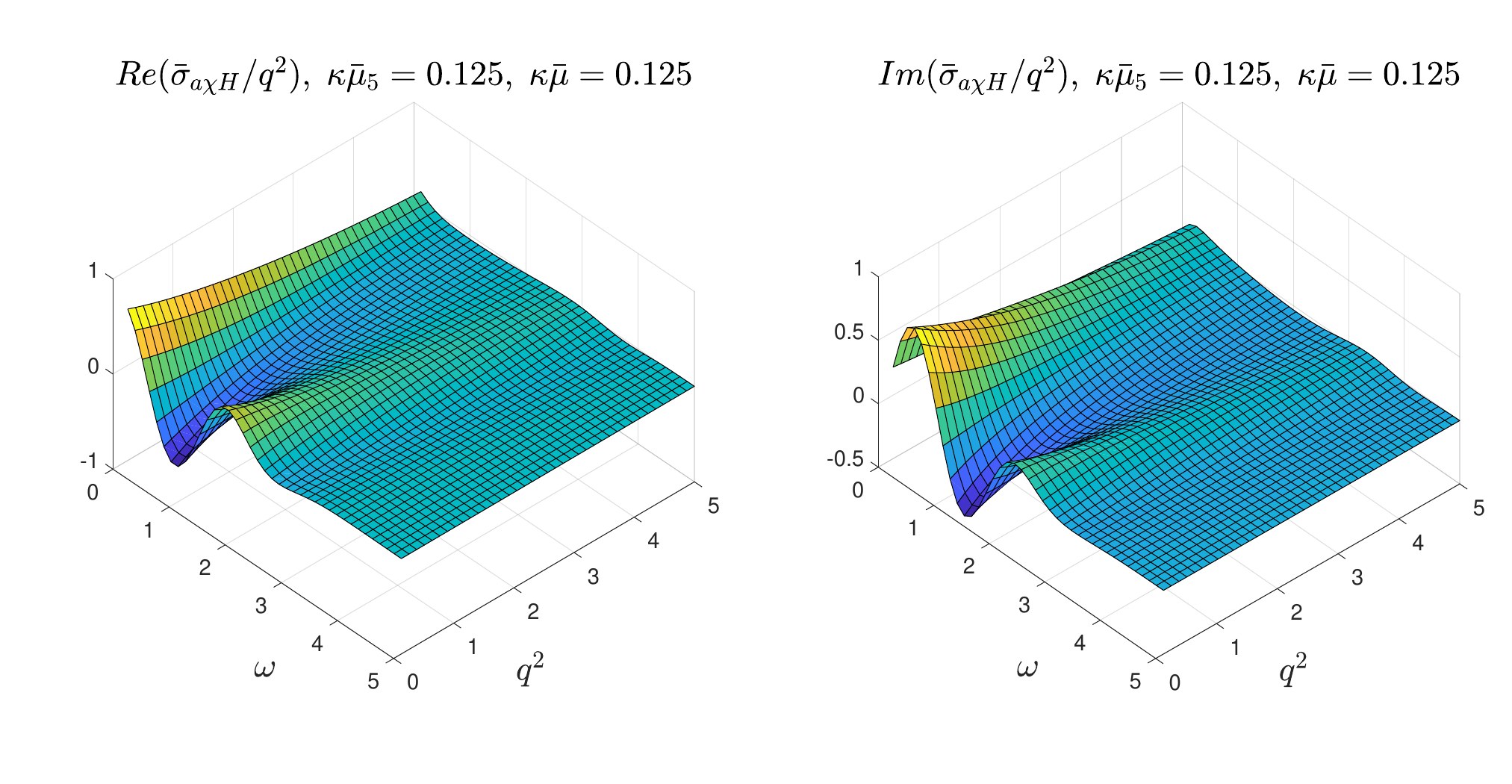}
\caption{TCF $\bar{\sigma}_{a\chi H}/q^2$ as a function of $\omega$ and $q^2$ when $\kappa\bar{\mu}=\kappa\bar{\mu}_5=1/8$.}\label{sigma_4xc}
\end{figure}

\begin{figure}[htbp]
\centering
\includegraphics[width=\textwidth]{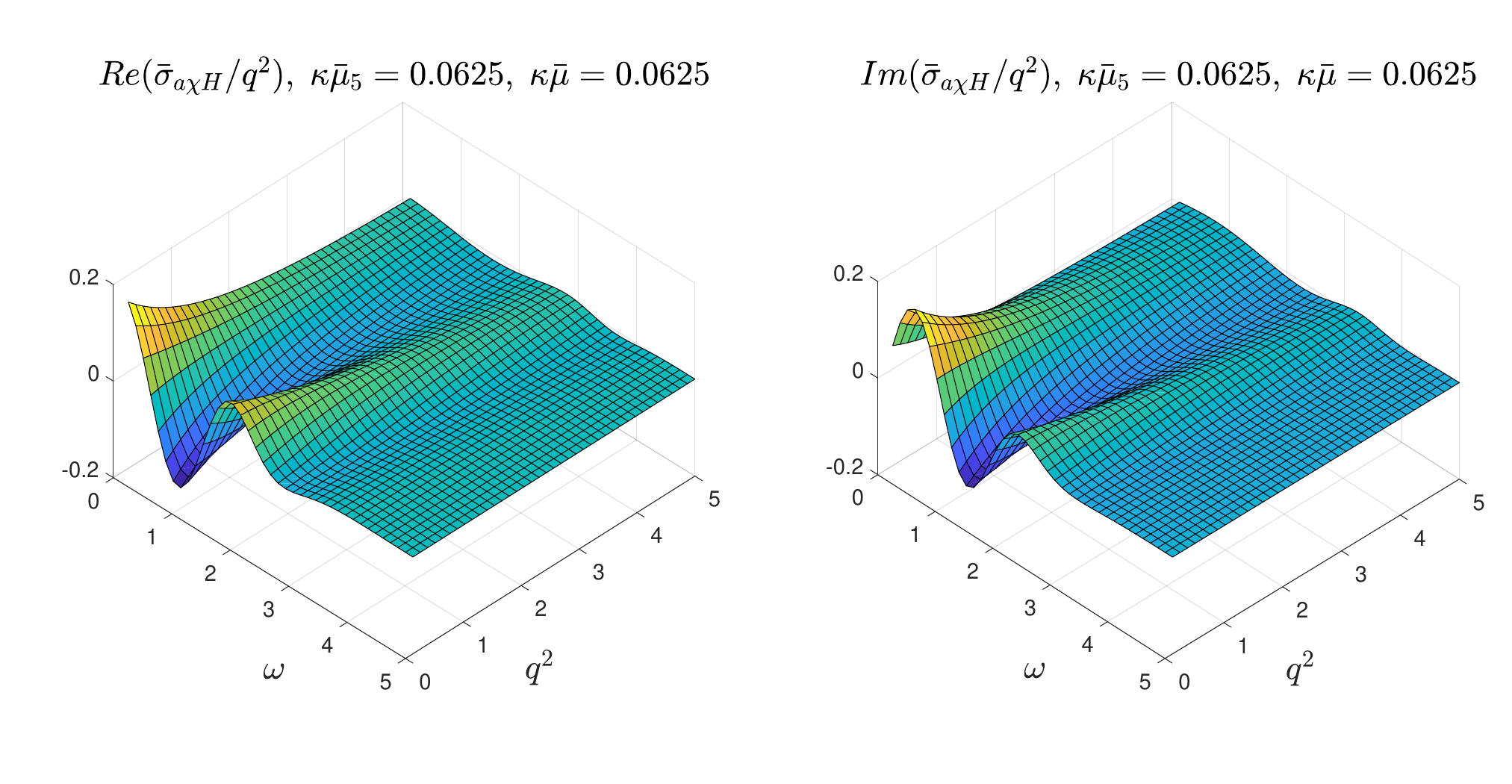}
\caption{TCF $\bar{\sigma}_{a\chi H}/q^2$ as a function of $\omega$ and $q^2$ when $\kappa\bar{\mu}=\kappa\bar{\mu}_5=1/16$.}\label{sigma_4xb}
\end{figure}

For representative values of $\bar{\mu},\bar{\mu}_5$, the frequency/momentum-dependence of TCFs $\mathcal{D}_H$, $\sigma_{a\chi H}$ and $\bar{\sigma}_{a\chi H}$ is displayed in Figures \ref{sigma_2xa}, \ref{sigma_2xb}, \ref{sigma_8xa}, \ref{sigma_8xa2}, \ref{sigma_4xc} and \ref{sigma_4xb}. These plots show similar behaviors as Figures \ref{sigma_chixa} and \ref{sigma_chixc}.
In contrast to $\sigma_{\bar{\chi}}$, the TCFs $\mathcal{D}_H$, $\sigma_{a\chi H}$ and  $\bar\sigma_{a\chi H}$ have non-trivial dependence on the chemical potentials for nonvanishing momentum values.

\begin{figure}[htbp]
\centering
\includegraphics[width=\textwidth]{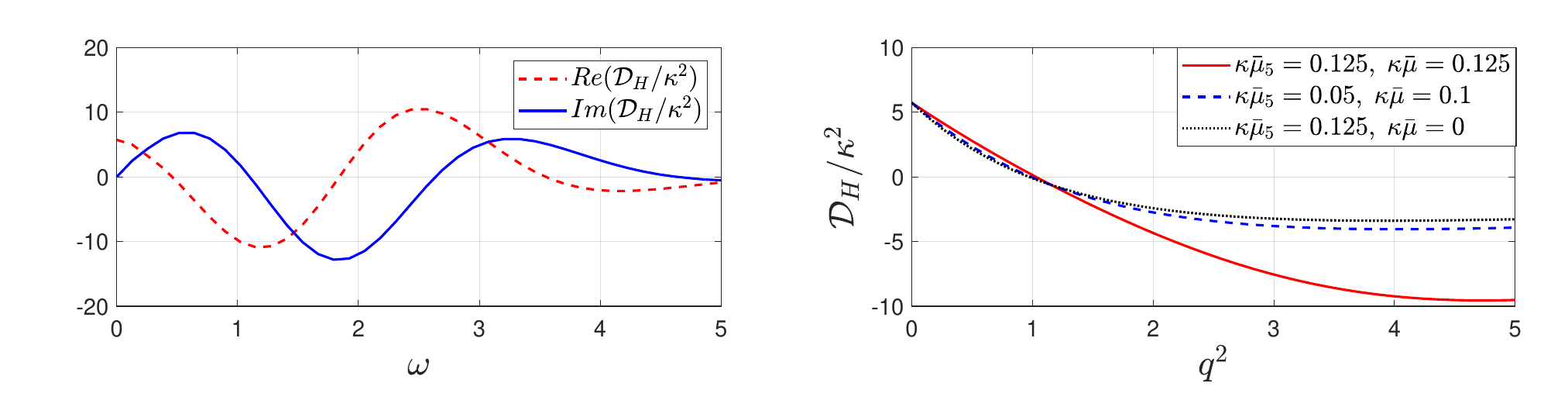}
\caption{$\omega$-dependence of $\mathcal{D}_H$ when $q=0$. Here $\mathcal{D}_H^0$ stands for DC limit of $\mathcal{D}_H$.}\label{sigma_2xc}
\end{figure}

\begin{figure}[htbp]
\centering
\includegraphics[width=\textwidth]{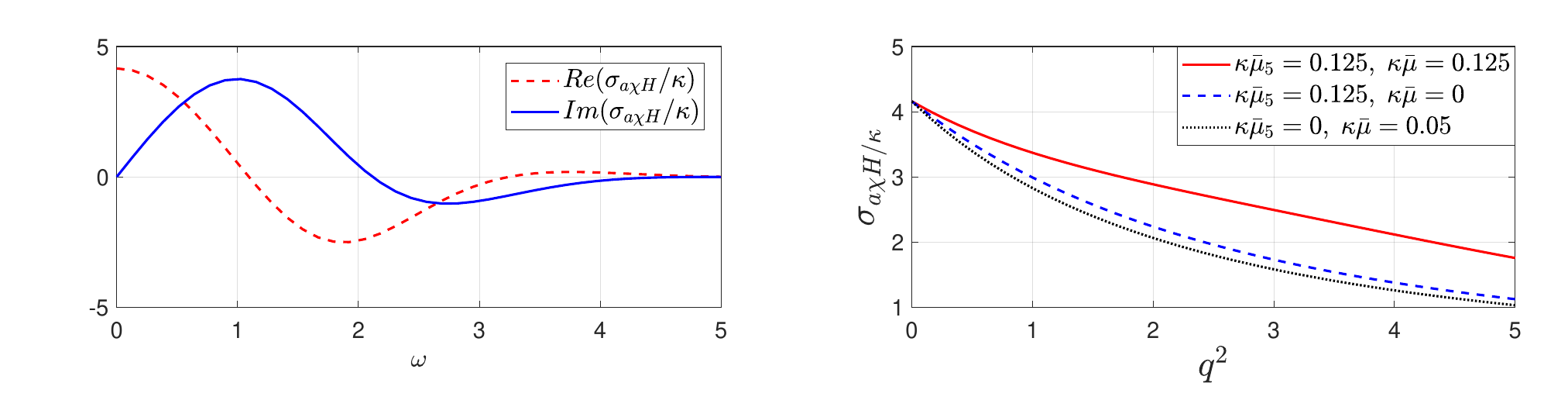}
\caption{$\omega$-dependence of $\sigma_{a\chi H}/\kappa $ when $q=0$ (left); $q^2$-dependence of $\sigma_{a\chi H} $ when $\omega=0$ (right). } \label{sigma_8xd}
\end{figure}

\begin{figure}[htbp]
\centering
\includegraphics[width=0.57\textwidth]{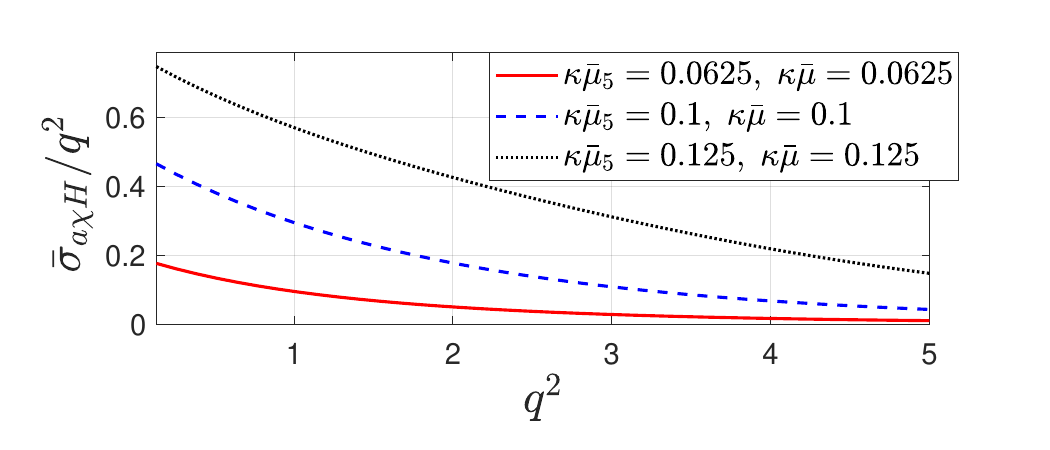}
\caption{$q^2$-dependence of $\bar{\sigma}_{a\chi H}$ when $\omega=0$. }
\label{sigma_4xd}
\end{figure}

Figures \ref{sigma_2xc}, \ref{sigma_8xd}, \ref{sigma_4xd}  display 2D slices of  \ref{sigma_2xa}, \ref{sigma_2xb}, \ref{sigma_8xa}, \ref{sigma_8xa2}, \ref{sigma_4xc} and \ref{sigma_4xb} when either $\omega=0$ or $q=0$. Recall that when $q=0$, $\mathcal{D}_H$ does not depend on chemical potentials, as can been checked from relevant ODEs (\ref{eom V1}, \ref{eom V3} ,\ref{eom Vb1}). Similarly, from ODEs (\ref{eom V4}, \ref{eom V6}, \ref{eom Vb4},\ref{ODEf}) it is obvious that when $q=0$, $\bar{\sigma}_{a\chi H}$ vanishes and $\sigma_{a\chi H}$ does not depend on chemical potentials. Once $q\neq 0$, TCFs $\mathcal{D}_H$, $\sigma_{a\chi H}$ and $\bar{\sigma}_{a\chi H}$ depend on chemical potentials non-linearly.



\begin{figure}[htbp]
\centering
\includegraphics[width=\textwidth]{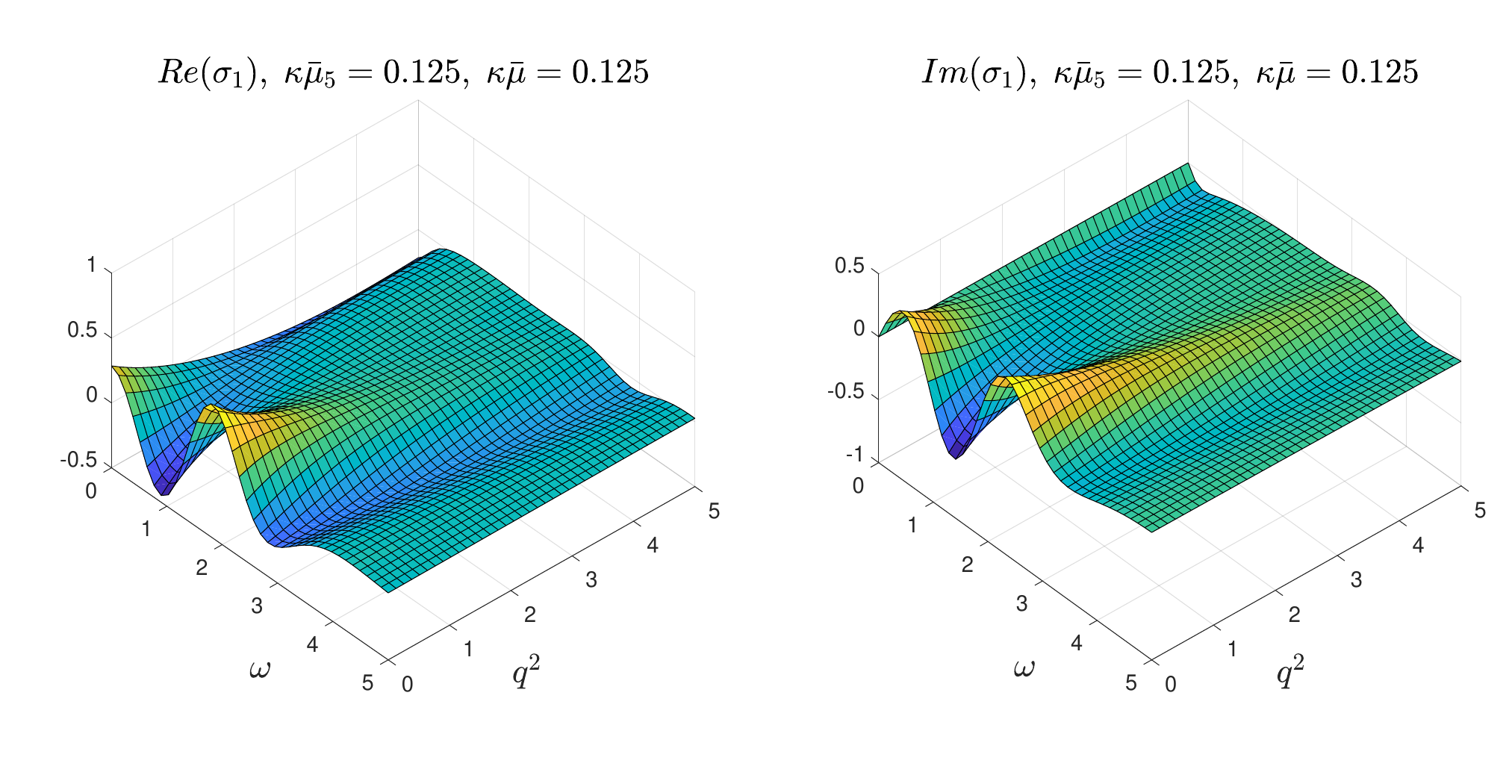}
\caption{TCF $\sigma_1$ as a function of $\omega$ and $q^2$ when $\kappa\bar{\mu}=\kappa\bar{\mu}_5=1/8$.}\label{sigma_1xa}
\end{figure}

\begin{figure}[htbp]
\centering
\includegraphics[width=\textwidth]{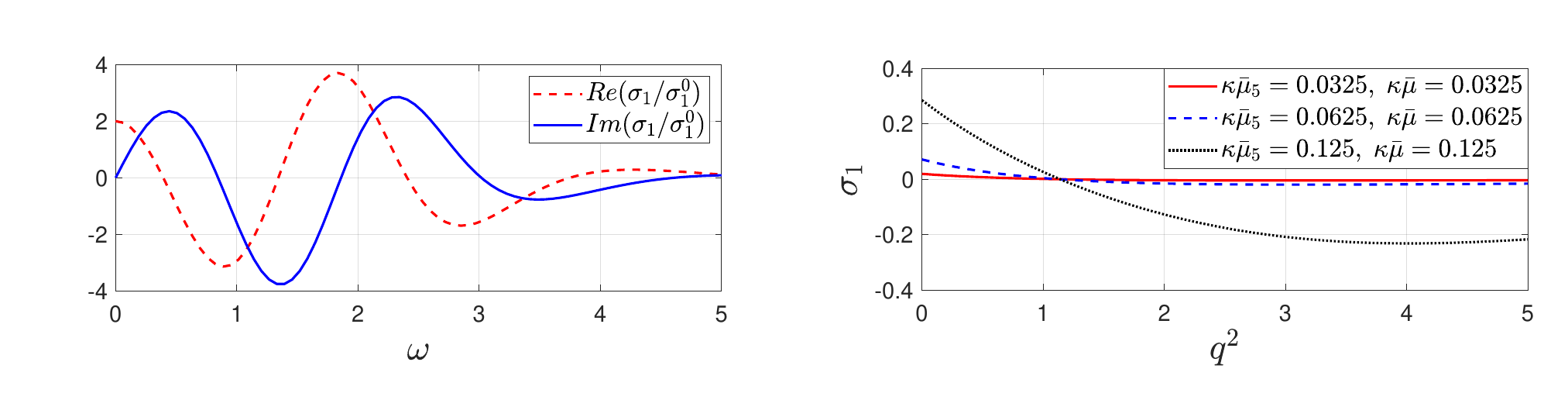}
\caption{$\omega$-dependence of $\sigma_{1}/\sigma_1^0 $ when $q=0$ (left); $q^2$-dependence of $\sigma_{1} $ when $\omega=0$ (right).}\label{sigma_1xc}
\end{figure}

\begin{figure}[htbp]
\centering
\includegraphics[width=\textwidth]{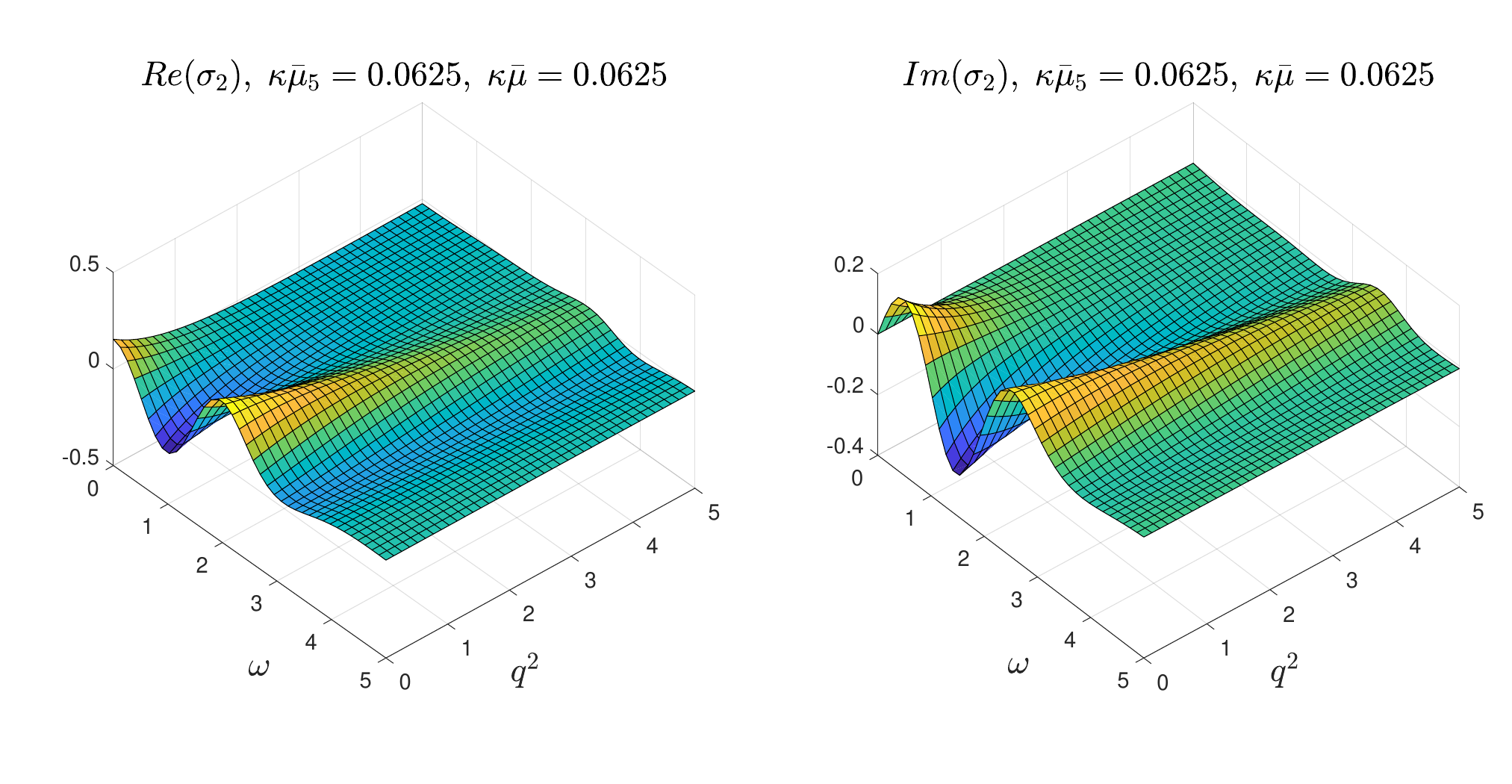}
\caption{TCF $\sigma_2$ as a function of $\omega$ and $q^2$ when $\kappa\bar{\mu}=\kappa\bar{\mu}_5=1/16$.}\label{sigma_5xa}
\end{figure}

\begin{figure}[htbp]
\centering
\includegraphics[width=\textwidth]{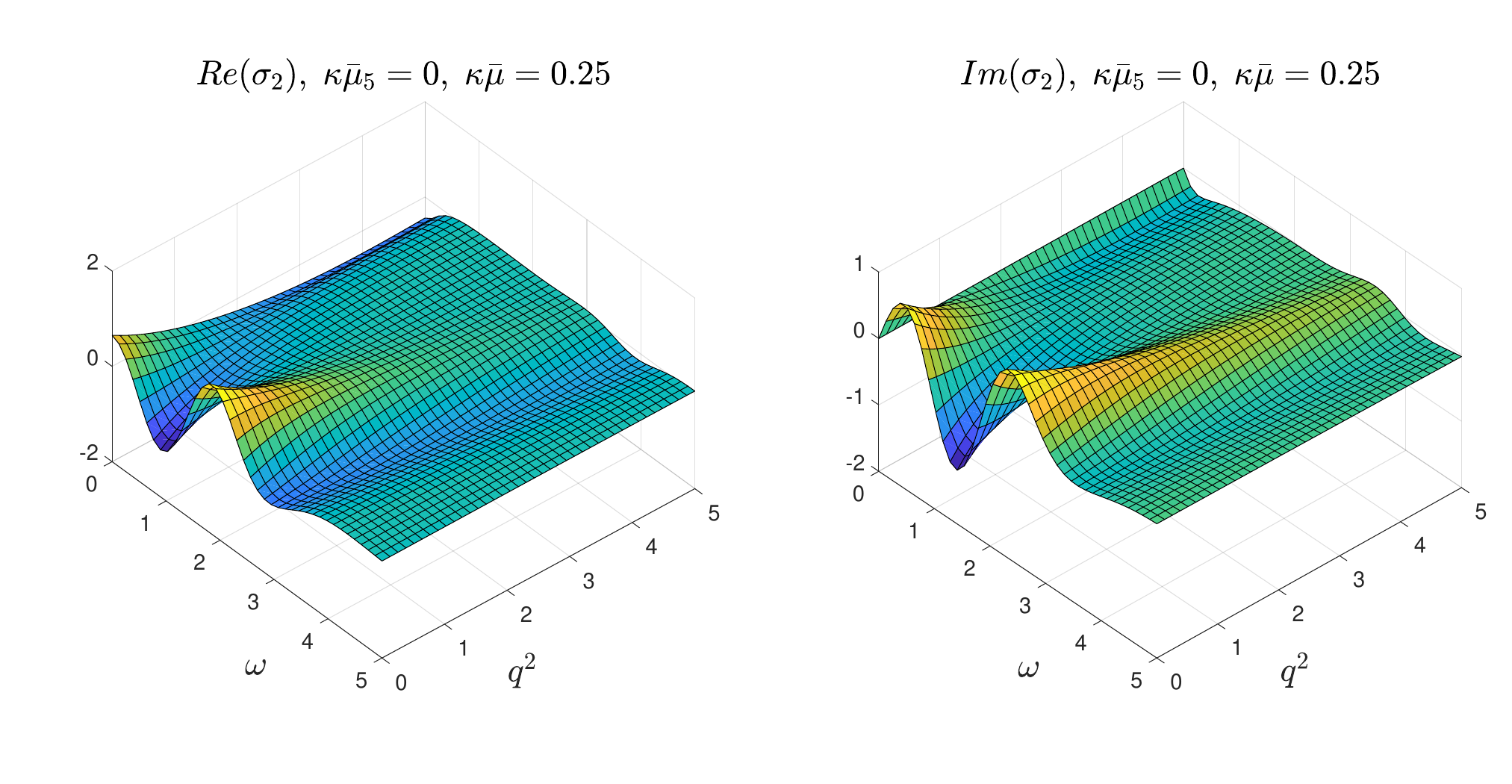}
\caption{TCF $\sigma_2$ as a function of $\omega$ and $q^2$ when $\kappa\bar{\mu}=1/4$, $\kappa\bar{\mu}_5=0$.}\label{sigma_5xb}
\end{figure}

\begin{figure}[htbp]
\centering
\includegraphics[width=\textwidth]{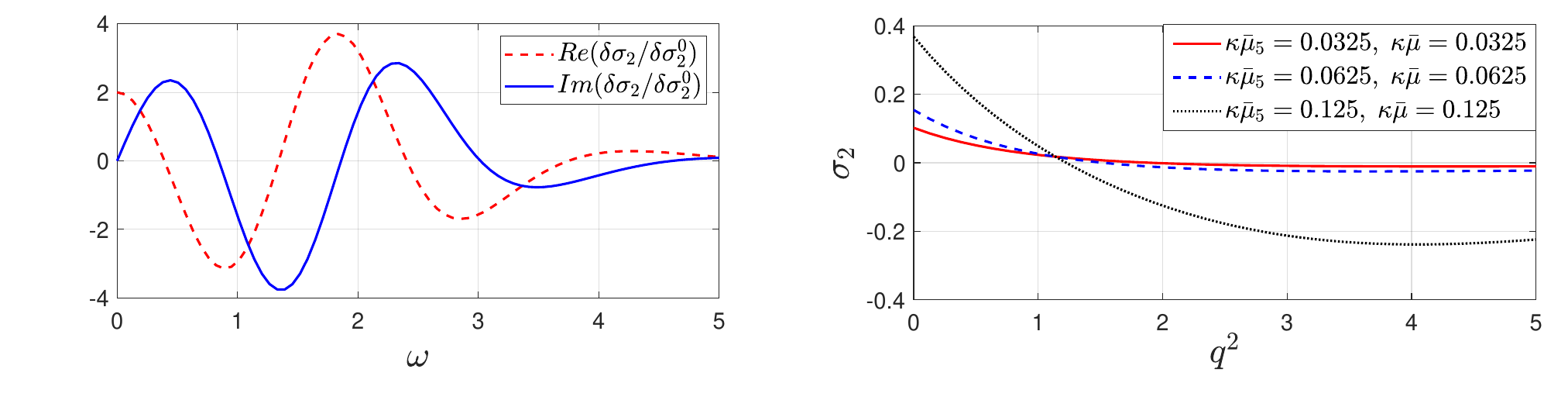}
\caption{$\omega$-dependence of $\delta\sigma_{2}/\delta\sigma_2^0 $ when $q=0$ (left); $q^2$-dependence of $\sigma_{2} $ when $\omega=0$ (right).}\label{sigma_22xc}
\end{figure}

\begin{figure}[htbp]
\centering
\includegraphics[width=\textwidth]{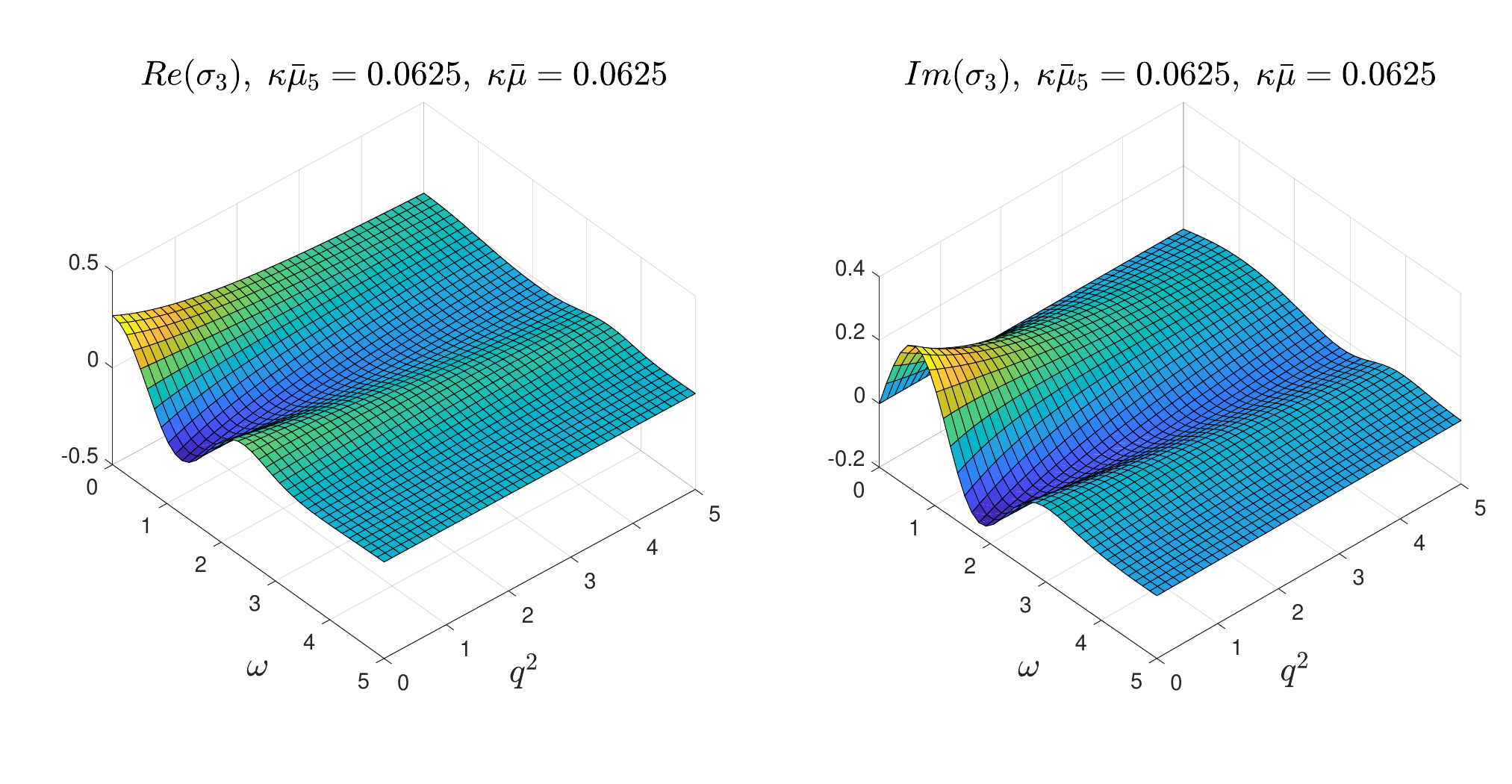}
\caption{Conductivity $\sigma_3$ as a function of $\omega$ and $q^2$ when $\kappa\bar{\mu}=\kappa\bar{\mu}_5=1/16$.}\label{sigma_3xa}
\end{figure}

\begin{figure}[htbp]
\centering
\includegraphics[width=\textwidth]{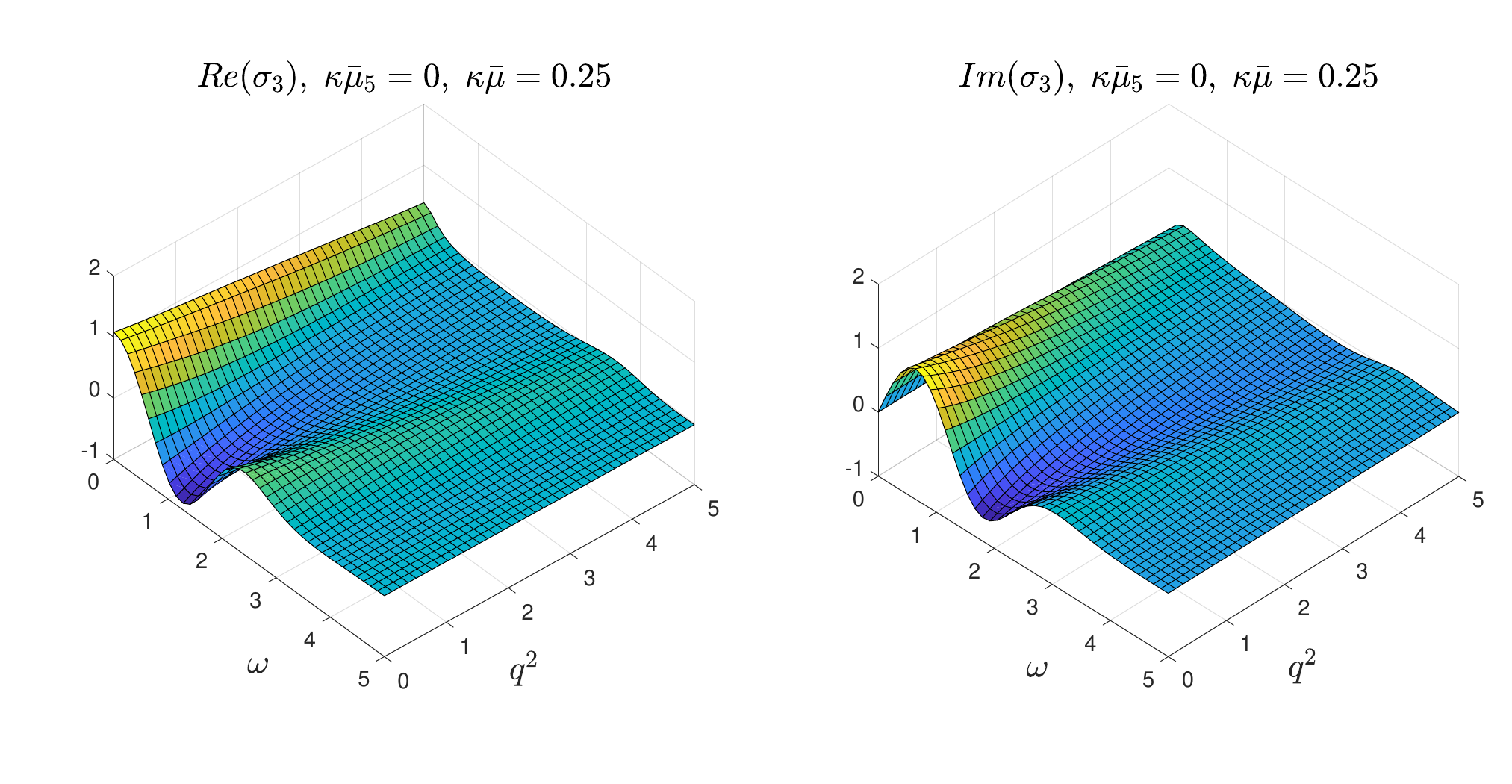}
\caption{Conductivity $\sigma_3$ as a function of $\omega$ and $q^2$ when $\kappa\bar{\mu}=1/4$, $\kappa\bar{\mu}_5=0$.}\label{sigma_3xb}
\end{figure}

\begin{figure}[htbp]
\centering
\includegraphics[width=\textwidth]{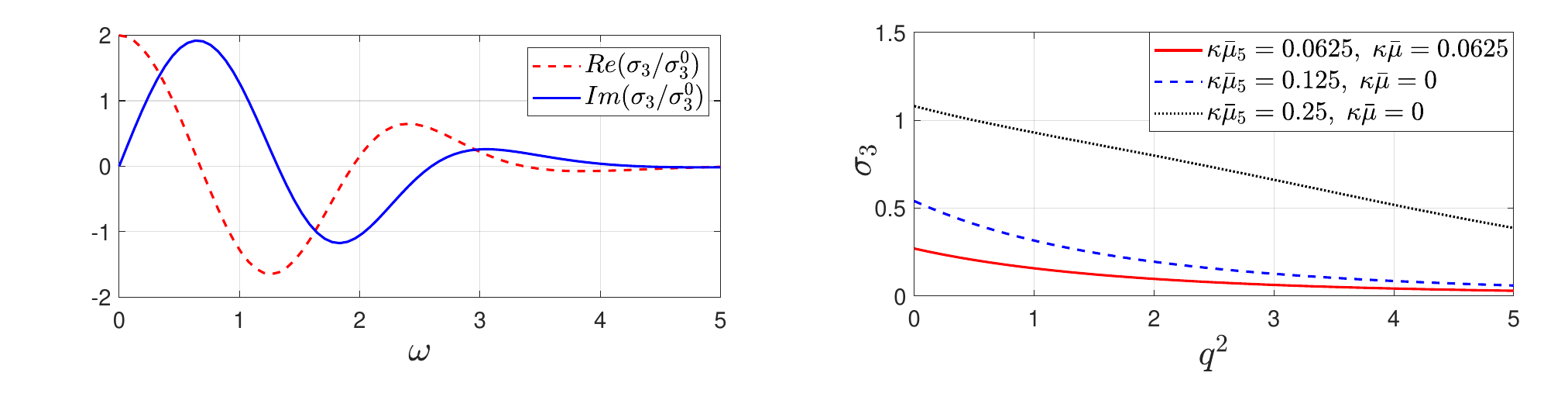}
\caption{$\omega$-dependence of $\sigma_{3}/\sigma_3^0 $ when $q=0$ (left); $q^2$-dependence of $\sigma_{3} $ when $\omega=0$ (right).}\label{sigma_3xc}
\end{figure}

Finally, we turn to the remaining TCFs $\sigma_{1,2,3}$ and $\bar{\sigma}_3$ which multiply third order derivative structures. The $\sigma_2$ and $\bar\sigma_3$ could be thought of as the axial analogues of $\sigma_1$ and $\sigma_3$, respectively. While $\sigma_2$ still has nonzero value when both $\kappa\bar{\mu}$ and $\kappa \bar{\mu}_5$ vanish, $\sigma_1$ relies on that $\kappa\bar{\mu}\bar{\mu}_5\neq 0$. Without loss of generality, we take $\kappa \bar{\mu} \geq \kappa \bar{\mu}_5$ when making plots for $\sigma_{1,2}$. Note that given the crossing rule (\ref{symmetric transports}), $\bar\sigma_3$ can be extracted from $\sigma_3$ by $\bar{\mu} \leftrightarrow \bar{\mu}_5$. For representative choices of $\bar{\mu},\bar{\mu}_5$, the 3D plots of these TCFs are summarised in Figures \ref{sigma_1xa}, \ref{sigma_5xa}, \ref{sigma_5xb}, \ref{sigma_3xa} and \ref{sigma_3xb}. In Figures  \ref{sigma_1xc}, \ref{sigma_22xc} and \ref{sigma_3xc} we depict 2D slices of Figures \ref{sigma_1xa}, \ref{sigma_5xa}, \ref{sigma_5xb}, \ref{sigma_3xa} and \ref{sigma_3xb} when either $q=0$ or $\omega=0$. As for $\mathcal{D}_H$, $\sigma_{a\chi H}$ and $\bar{\sigma}_{a\chi H}$, for nonzero $q$, $\sigma_{1,2,3}$ and $\bar{\sigma}_3$ depend on chemical potentials non-linearly.

The universal dependence on vector/axial potentials at $q=0$ is revealed by considering the normalized quantities $\sigma_1/\sigma_1^0$, $\sigma_3/\sigma_3^0$, $\delta\sigma_2/\delta\sigma_2^0$. Here  $\sigma_1^0$, $\sigma_3^0$, $\delta\sigma_2^0$ stands for DC limit of the corresponding TCFs and $\delta\sigma_2=\sigma_2-\sigma_2(\kappa\bar{\mu}=\kappa\bar{\mu}_5=0)$. As seen from (\ref{eom V2}) and (\ref{eom Vb2}), $\sigma_1/\sigma_1^0$ and $\delta\sigma_2/\delta\sigma_2^0$ are identical at $q=0$. Thus, we will mainly focus on $\sigma_1/\sigma_1^0$. $\omega$-dependence of $\sigma_1/\sigma_1^0$ and $\sigma_3/\sigma_3^0$ is displayed in Figures \ref{sigma_1xc} and \ref{sigma_3xc}. We observe the universal dependence of vector/axial potentials at $q=0$, that is to say these normalised quantities do not depend on chemical potentials. Explicitly, $\sigma_1$ is linear in $\kappa^2\bar{\mu}\bar{\mu}_5$. $\sigma_3$ is linear in $\kappa\bar{\mu}$. $\sigma_2$ has anomalous correction which is linear in $\kappa^2(\bar{\mu}^2+\bar{\mu}_5^2)$. All these features can also be realised from the corresponding ODEs. Note that by employing crossing rule (\ref{symmetric transports}), $\bar\sigma_3$ is linear in $\kappa\bar{\mu}_5$.




\subsection{CMW dispersion relation to all orders: non-dissipative modes} \label{s44}

The TCF $\sigma_{\bar{\chi}}$ enters the dispersion relation of CMW:
\begin{equation}\label{CMW1}
\omega=\pm\sigma_{\bar{\chi}}(\omega,q^2) \,\kappa\vec q\cdot\vec {\bf B} - i \mathcal{D}(\omega,q^2)  q^2.
\end{equation}
The dispersion relation (\ref{CMW1}) is exact to all orders in $q^2$, provided $\kappa{\bf  B}\ll 1$.
General solutions of this equation are complex and cannot be studied with our present results. This is because  $ \sigma_{\bar{\chi}}(\omega,q^2)$ and $\mathcal{D}(\omega,q^2) $ have been computed
for real values of $\omega$ only.   We believe that
beyond the hydrodynamic limit, equation (\ref{CMW}) has infinitely many gapped modes. Exploring this point in general would require going
into complex $\omega$ plane for the TCFs, which is beyond the scope of the present work.  Yet, quite intriguingly, there is a set of purely real non-dissipative solutions to (\ref{CMW1}).
In order to find these solutions we have devised the following procedure.

\begin{figure}[htbp]
\centering
\includegraphics[width=\textwidth]{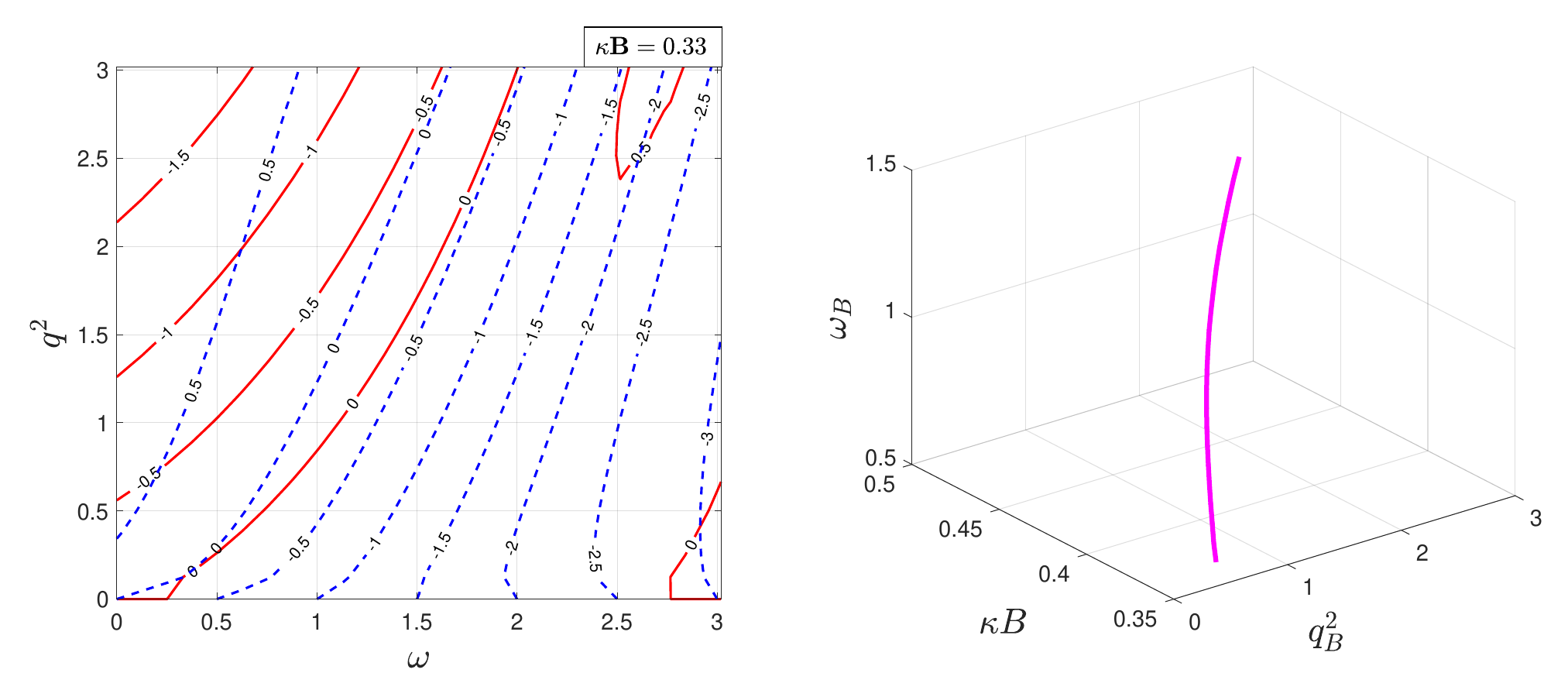}
\caption{Contour plots for the functions $\phi_R$ (blue dashed) and $\phi_I$ (red solid) at $\kappa {\bf B}=0.33$ (left); continuum of discrete non-dissipative
modes  ($\omega_B, q_B$) as function of $\kappa {\bf B}$ (right). }\label{ab}
\end{figure}

First, the equation is split into real and imaginary parts (assuming $\vec q$ parallel to $\vec{\bf B}$):
\begin{eqnarray}
\phi_I(\omega,q^2, \kappa{\bf B})&&\equiv \rm Im[\sigma_{\bar{\chi}}(\omega,q^2)] \,\kappa q {\bf B} - Re[ \mathcal{D}(\omega,q^2)]\,  q^2, \nonumber\\
\phi_R(\omega,q^2, \kappa{\bf B})&&\equiv -\omega+ \rm Re[\sigma_{\bar{\chi}}(\omega,q^2)] \,\kappa q {\bf B} +  Im[\mathcal{D}(\omega,q^2)]\,  q^2.
\end{eqnarray}
For a fixed value of $\kappa{\bf B}$, say $\kappa{\bf B}=0.33$, the functions $\phi_I$ and $\phi_R$ are shown in Figure \ref{ab} (left) as contour plots in $(\omega, q^2)$ space
(the function $ \mathcal{D}(\omega,q^2) $  is taken from \cite{1511.08789} ).
The dashed (blue) and solid (red) curves stand for $\phi_I$ and $\phi_R$ respectively. The numbers indicated on the curves correspond to the values of these functions along the curves.
Our interest is when both functions vanish simultaneously, that is a crossing point of $\phi_I=0$ and $\phi_R=0$ curves.  Such crossing is clearly seen in the region $\omega<0.5$ and $q^2<0.5$.
We denote this point by ($\omega_B, q_B$).  This is a discrete density wave mode propagating in the medium without any dissipation.

The procedure could be repeated for other values of $\kappa {\bf  B}$. The result is a one dimensional curve in a 3d parameter space depicted in Fig. \ref{ab} (right).
A few comments are in order. First, there is a minimal value of $\kappa {\bf B}\simeq 0.33$ for which there exists such a solution. Second, in fact there are multiple solutions corresponding to several disconnected
branches in Figure \ref{ab}, which we do not display.

\section{Conclusion}\label{s5}

In this work, we have continued exploration of nonlinear chiral anomaly-induced transport
phenomena based on a holographic model with two $U(1)$ fields interacting via gauge Chern-Simons terms. For a  finite temperature system, we constructed off-shell constitutive relations
for the vector and axial currents.

The constitutive relations contain nine terms which are  linear simultaneously in the charge density fluctuations and  constant background external fields.
The nine terms  summarised in (\ref{ji11},\ref{j5i11}) correspond to all order  resummation of gradients of the the charge density fluctuations parameterised by TCFs,
 first computed analytically in the hydrodynamic limit (section \ref{s42}) and then numerically for large frequency/momentum (section \ref{s43}).
 A common feature of all TCFs in (\ref{ji11},\ref{j5i11}) is that they depend  weakly on spatial momentum  but display  pronounced dependence
 on frequency in the form of damped oscillations  vanishing asymptotically at $\omega\simeq 5$.

Most of our results are presented in Summary section \ref{s2}.  Among new results worth highlighting is the CME memory function computation $\tilde \sigma_{\bar\chi}(t-t^\prime)$.
The memory function is found to differ dramatically from a delta-function form of instantaneous response.  In fact,  $\tilde \sigma_{\bar\chi}(t-t^\prime)$ vanishes at $t=t^\prime$ and
the CME response gets built only after a finite amount of time of order temperature.

Another result we find of interest is related to CMW dispersion relation, which for the first time  was considered  to all orders in momentum $q$.  Beyond the perturbative hydrodynamic limit,
we found a continuum set of discrete density wave modes, which can propagate in the medium without any dissipation. While the original CMW dissipates and that could be one of the problems
for its  detection, the new modes that we discover should be long lived and have  some potential experimental signature\footnote{Obviously, if an experimentally accessible chiral plasma shares
similar features as discovered within our holographic model.}.  It is important to remember that our calculation of the CMW dispersion
relation is done for a weak magnetic field only.  One can obviously question the validity of the results beyond this approximation.  Both
TCFs $ \sigma_{\bar{\chi}}$ and $\mathcal{D}$ that enter the CMW dispersion relation are   functions of  $\vec{\bf E}$ and $\vec{\bf B}$. In  our previous work  \cite{Bu:2018psl}, we initiated
this study, still in perturbative  in $\vec{\bf E}$ and $\vec{\bf B}$ regions,  but a full non-perturbative analysis will be reported elsewhere \cite{Part3}.

We have found a wealth of non-linear phenomena all induced entirely by the chiral anomaly. An important next step in deriving a full chiral MHD would be to abandon
the probe limit adopted in this paper and include the dynamics of a neutral flow as well.  This will bring into the picture additional effects such as thermoelectric conductivities, normal Hall current,
the chiral vortical effect \cite{0809.2488,0809.2596}, and some nonlinear effects discussed in \cite{1603.03620}.  We plan to address these in the future.

\appendix
\section{Supplement for section \ref{s4}}
\subsection{ODEs and the constraints for the decomposition coefficients in (\ref{decomposition Vt11}-\ref{decomposition Ai11})}\label{appendixb2}

We first collect the ODEs satisfied by the decomposition coefficients in (\ref{decomposition Vt11}-\ref{decomposition Ai11}) and then derive some constraint relations obeyed by these coefficients. Plugging (\ref{decomposition Vt11}-\ref{decomposition Ai11}) into (\ref{Vt(11)}-\ref{Ai(11)}) and performing Fourier transform $\partial_{\mu}\rightarrow(-i\omega,i\vec q)$, we obtained ODEs for the decomposition coefficients $S_i,\bar{S}_i,V_i,\bar{V}_i$. These ODEs can be grouped into partially decoupled sub-sectors:
\\
\underline{sub-sector (i):} $\{S_1,\bar{S}_1,V_1,\bar{V_1},V_2,\bar{V_2},V_3,\bar{V_3}\}$
\begin{equation}\label{ODEb}
0=r^2 \partial_r^2 S_1 + 3r \partial_r S_1 + \partial_r (V_1 -q^2V_2),
\end{equation}
\begin{equation} \label{eom Sb1}
0=r^2 \partial_r^2 \bar{S}_1 + 3r \partial_r \bar{S}_1 + \partial_r (\bar{V}_1 -q^2 \bar{V}_2) +\frac{12}{r} \partial_r g_4,
\end{equation}
\begin{equation} \label{eom V1}
0=(r^5-r)\partial_r^2 V_1 +(3r^4+1-2i\omega r^3)\partial_r V_1 -(i\omega r^2+q^2 r) V_1- \frac{12q^2}{r}\kappa(\bar{\rho}_5 V_3+\bar{\rho} \bar{V}_3),
\end{equation}
\begin{equation}\label{eom V2}
\begin{split}
0&=(r^5-r)\partial_r^2 V_2 +(3r^4+1-2i\omega r^3)\partial_r V_2 -i\omega r^2 V_2 -r V_1 -r^2(S_1+r\partial_r S_1)\\
&-\frac{12\kappa}{r}(\bar{\rho}_5 V_3+ \bar{\rho} \bar{V}_3),
\end{split}
\end{equation}
\begin{equation} \label{eom V3}
\begin{split}
0&=(r^5-r)\partial_r^2 V_3 +(3r^4+1-2i\omega r^3)\partial_r V_3 -(i\omega r^2+q^2 r) V_3- \frac{12\kappa}{r}(\bar{\rho}_5 V_1+\bar{\rho} \bar{V_1})\\
&+12\kappa \bar{\rho}r^2\partial_r f_2(i\omega g_4+g_3),
\end{split}
\end{equation}
\begin{equation} \label{eom Vb1}
\begin{split}
0&=(r^5-r)\partial_r^2 \bar{V}_1 +(3r^4+1-2i\omega r^3)\partial_r \bar{V}_1 -(i\omega r^2+q^2 r) \bar{V}_1- \frac{12q^2}{r}\kappa(\bar{\rho} V_3+\bar{\rho}_5 \bar{V}_3)\\
&+\frac{12}{r}(1+r^3\partial_r g_3),
\end{split}
\end{equation}
\begin{equation}\label{eom Vb2}
\begin{split}
0&=(r^5-r)\partial_r^2 \bar{V}_2 +(3r^4+1-2i\omega r^3)\partial_r \bar{V}_2 -i\omega r^2\bar{V}_2- r\bar{V}_1-r^2(\bar{S}_1+r\partial_r \bar{S}_1) \\ &-\frac{12\kappa}{r}(\bar{\rho} V_3+\bar{\rho}_5 \bar{V}_3),
\end{split}
\end{equation}
\begin{equation}
\begin{split}
0&=(r^5-r)\partial_r^2 \bar{V}_3 +(3r^4+1-2i\omega r^3)\partial_r \bar{V}_3 -(i\omega r^2+q^2 r)\bar{V}_3-\frac{12\kappa}{r}(\bar{\rho} V_1+\bar{\rho}_5 \bar{V_1})\\
&+12\kappa \bar{\rho}_5 r^2\partial_r f_2 (i\omega g_4+g_3)-6\kappa \bar{\rho}_5 \partial_r f_2.
\end{split}
\end{equation}
\\
\underline{sub-sector (ii):} $\{S_2,\bar{S}_2,V_4,\bar{V_4},V_5,\bar{V_5},V_6,\bar{V_6}\}$
\begin{equation}
0=r^2 \partial_r^2 S_2 + 3r \partial_r S_2 + \partial_r (V_4 -q^2V_5),
\end{equation}
\begin{equation}
0=r^2 \partial_r^2 \bar{S}_2 + 3r \partial_r \bar{S}_2 + \partial_r (\bar{V}_4 -q^2\bar{V}_5),
\end{equation}
\begin{equation} \label{eom V4}
0=(r^5-r)\partial_r^2 V_4 +(3r^4+1-2i\omega r^3)\partial_r V_4 -(i\omega r^2+q^2 r) V_4- \frac{12q^2}{r}\kappa(\bar{\rho}_5 V_6+\bar{\rho} \bar{V}_6),
\end{equation}
\begin{equation}
\begin{split}
0&=(r^5-r)\partial_r^2 V_5 +(3r^4+1-2i\omega r^3)\partial_r V_5 -i\omega r^2V_5-rV_4 -r^2 (S_2+r\partial_r S_2)\\
&-\frac{12\kappa}{r}(\bar{\rho}_5 V_6+\bar{\rho} \bar{V}_6),
\end{split}
\end{equation}
\begin{equation} \label{eom V6}
0=(r^5-r)\partial_r^2 V_6 +(3r^4+1-2i\omega r^3)\partial_r V_6 -(i\omega r^2+q^2 r)V_6-\frac{12\kappa}{r}(\bar{\rho}_5 V_4+\bar{\rho} \bar{V_4}),
\end{equation}
\begin{equation} \label{eom Vb4}
0=(r^5-r)\partial_r^2 \bar{V}_4 +(3r^4+1-2i\omega r^3)\partial_r \bar{V}_4 -(i\omega r^2 +q^2 r) \bar{V}_4- \frac{12q^2}{r}\kappa(\bar{\rho} V_6+\bar{\rho}_5 \bar{V}_6),
\end{equation}
\begin{equation}
\begin{split}
0&=(r^5-r)\partial_r^2 \bar{V}_5 +(3r^4+1-2i\omega r^3)\partial_r \bar{V}_5 -i\omega r^2 \bar{V}_5-r \bar{V}_4-r^2(\bar{S}_2+r\partial_r \bar{S}_2)\\
&-\frac{12\kappa}{r}(\bar{\rho} V_6+\bar{\rho}_5 \bar{V}_6),
\end{split}
\end{equation}
\begin{equation}\label{ODEf}
\begin{split}
0&=(r^5-r)\partial_r^2 \bar{V}_6 +(3r^4+1-2i\omega r^3)\partial_r \bar{V}_6 -(i\omega r^2 +q^2 r)\bar{V}_6-\frac{12\kappa}{r}(\bar{\rho} V_4+\bar{\rho}_5 \bar{V}_4)\\
&-12 r^2 \left[\partial_r g_4-\partial_r f_1 (i\omega g_4+g_3)\right]-6 \partial_r f_1.
\end{split}
\end{equation}

The remaining decomposition coefficients satisfy the same ODEs as above. More specifically, the sub-sector $\{\bar{S}_3,S_3,\bar{V}_7,V_7,\bar{V}_8,V_8,\bar{V}_9,V_9\}$ satisfies the same equations as the sub-sector (i): $\{S_1,\bar{S}_1,V_1,\bar{V_1},V_2,\bar{V_2},V_3,\bar{V_3}\}$; the sub-sector $\{\bar{S}_4,S_4,\bar{V}_{10},V_{10},\bar{V_{11}},V_{11},\bar{V}_{12},V_{12}\}$ obeys the same equations as sub-sector (ii): $\{S_2,\bar{S}_2,V_4,\bar{V_4},V_5,\bar{V_5},V_6,\bar{V_6}\}$.

In what follows, we explore some ``mirror symmetry relations" among these decomposition coefficients, which are useful in simplifying the expressions for currents' constitutive relations at the order $\mathcal{O}(\epsilon^1\alpha^1)$. First, notice that
\begin{equation} \label{ident1}
\{S_3,\bar{S}_3,V_7,\bar{V_7},V_8,\bar{V_8},V_9,\bar{V_9}\}= \{\bar{S}_1,S_1,\bar{V}_1,V_1,\bar{V}_2,V_2,\bar{V}_3,V_3\},
\end{equation}
since these two sub-sectors satisfy identical system of ODEs and have the same boundary conditions. Following this reasoning,
\begin{equation} \label{ident2}
\{S_4,\bar{S}_4,V_{10},\bar{V}_{10},V_{11},\bar{V_{11}}, V_{12}, \bar{V_{12}}\} = \{\bar{S}_2,S_2,\bar{V}_4,V_4,\bar{V}_5,V_5,\bar{V}_6,V_6\}.
\end{equation}
The ``equal sign" in (\ref{ident1},\ref{ident2}) should be understood in the specific order as shown therein.

Certain relations can be established among the decomposition coefficients in (\ref{decomposition Vt11}-\ref{decomposition Ai11}).
It follows from the  Landau frame convention (\ref{si=0}) and  boundary conditions (\ref{bc1},\ref{bc2}) that
\begin{equation}\label{constraints}
\begin{split}
&S_1=0, \qquad V_1-q^2V_2=0, \qquad S_2=0, \qquad V_4-q^2V_5=0\\
&\bar{S}_2=0, \qquad \qquad \bar{V}_4-q^2\bar{V}_5=0.
\end{split}
\end{equation}

Now lets explore the mirror symmetry for the decomposition coefficients under exchange  $\bar\rho \leftrightarrow \bar \rho_5$. The decomposition coefficients in the sub-sector $\{S_1,\bar{S}_1,V_1,\bar{V_1},V_2,\bar{V_2},V_3,\bar{V_3}\}$ are found symmetric with respect to $\bar\rho,\bar\rho_5$:
\begin{equation} \label{symmetric1}
\begin{split}
&V_i(\bar\rho,\bar\rho_5)= V_i(\bar\rho_5,\bar\rho),\qquad \bar V_i(\bar\rho,\bar\rho_5)= \bar V_i(\bar\rho_5,\bar\rho),\qquad i=1,2,3,\\ &S_1(\bar\rho,\bar\rho_5)= S_1(\bar\rho_5,\bar\rho),\qquad
\bar S_1(\bar\rho,\bar\rho_5)= \bar S_1(\bar\rho_5,\bar\rho).
\end{split}
\end{equation}
Similarly, in the second sub-sector $\{S_2,\bar{S}_2,V_4,\bar{V_4},V_5,\bar{V_5},V_6,\bar{V_6}\}$,
\begin{equation} \label{symmetric2}
\begin{split}
&\bar S_2(\bar\rho, \bar\rho_5)=S_2(\bar\rho_5, \bar\rho),\qquad \bar V_i(\bar\rho, \bar\rho_5)=V_i(\bar\rho_5, \bar\rho),\qquad i=4,5\\
&V_6(\bar\rho,\bar\rho_5)=V_6(\bar\rho_5,\bar\rho),\qquad \bar V_6(\bar\rho,\bar\rho_5) = \bar V_6(\bar\rho_5,\bar\rho).
\end{split}
\end{equation}

The symmetry relations (\ref{symmetric1},\ref{symmetric2}) guide the choice of values for $\kappa\bar \rho$ and $\kappa\bar\rho_5$ in numerical procedure for the ODEs. Given these relations, the choice $\kappa\bar \rho\geqslant \kappa \bar \rho_5$  can be applied when solving $\left\{V_1,\bar V_1, V_3, \bar V_3\right\}$, $\left\{S_1,V_2,\bar{S}_1,\bar V_2\right\}$ and $\left\{V_4,\bar V_4,V_6, \bar V_6\right\}$ without loosing generality. These relations  also help to reduce the number of the ODEs to be solved.

\subsection{Perturbative solutions}\label{pertubativs}

Here, we summarise the perturbative solutions of (\ref{ODEb}-\ref{ODEf}) in the hydrodynamic limit $\omega,q\ll1$. Recall that the decomposition coefficients are formally expanded as (\ref{pertdec}). Then, at each order in the hydrodynamic expansion, solutions are expressed as double integrals over $r$. The final results, up to third order in derivative expansion, are listed below.\\
sub-sector (i): $\left\{ S_1,\bar{S}_1, V_1,\bar{V}_1,V_2,\bar{V}_2,V_3,\bar{V}_3 \right\}$:
\begin{equation}\label{pertb}
S_1^{(0)}=V_1^{(0)}=0,
\end{equation}
\begin{equation}
\bar{V}_{1}^{(0)}=\int^{\infty}_{r} \frac{xdx}{x^4-1}\int^{x}_1 dy \frac{12}{y^3} =3 \log \frac{1+r^2}{r^2}\xrightarrow[]{ r\rightarrow\infty } \frac{3}{r^2}+\mathcal{O}\left(\frac{1}{r^3}\right),
\end{equation}
\begin{equation}
\bar{S}_{1}^{(0)}=-\int^{\infty}_{r} \frac{dx}{x^3}\int^{\infty}_x dy\left[y \partial_y \bar{V}_1^{(0)}+12 \partial_y g_4^{(0)}\right]\xrightarrow[]{ r\rightarrow\infty } \mathcal{O}\left(\frac{1}{r^3}\right),
\end{equation}
\begin{equation}
V_{3}^{(0)}=-\int^{\infty}_{r} \frac{xdx}{x^4-1}\int^{x}_1 dy \frac{6 \kappa}{y^3} \bar{\rho}\left[2 \bar{V}_1^{(0)}+y \partial_y f_2 \right]\xrightarrow[]{ r \rightarrow \infty } \frac{9\kappa \bar{\rho}}{r^2}(2-3 \log 2) +\mathcal{O} \left(\frac{1}{r^3} \right),
\end{equation}
\begin{equation}
\bar{V}_{3}^{(0)}=-\int^{\infty}_{r} \frac{xdx}{x^4-1}\int^{x}_1 dy \frac{6 \kappa}{y^3} \bar{\rho}_5\left[2 \bar{V}_1^{(0)}+y \partial_y f_2 \right]\xrightarrow[]{ r \rightarrow \infty } \frac{9\kappa \bar{\rho}_5}{r^2}(2-3 \log 2)+\mathcal{O} \left(\frac{1}{r^3}\right),
\end{equation}
\begin{equation}
\begin{split}
V_{2}^{(0)}&=-\int^{\infty}_{r} \frac{xdx}{x^4-1}\int^{x}_1 dy \frac{12 \kappa}{y^3}(\bar{\rho}_5 V_3^{(0)}+\bar{\rho} \bar{V}_3^{(0)})\\
&\xrightarrow[]{ r\rightarrow\infty }\frac{27 \kappa^2 \bar{\rho} \bar{\rho}_5} {r^2}[6+\log 2(5\log 2 -12)]+\mathcal{O}\left(\frac{1}{r^3}\right),
\end{split}
\end{equation}
\begin{eqnarray}
&&\bar{V}_{2}^{(0)}=-\int^{\infty}_{r} \frac{xdx}{x^4-1}\int^{x}_1 dy \left[\bar{S}_1^{(0)}+y \partial_y\bar{S}_1^{(0)}+\frac{1}{y} \bar{V}_1^{(0)}+\frac{12 \kappa}{y^3}(\bar{\rho} V_3^{(0)}+\bar{\rho}_5 \bar{V}_3^{(0)})\right]\\
&&\qquad \quad\xrightarrow[]{ r\rightarrow\infty }\frac{1}{2r^2}\left\lbrace \frac{1}{8}(6\pi-\pi^2-12\log2)+27 \kappa^2 (\bar{\rho}^2+\bar{\rho}_5^2)[6+\log 2(5 \log 2 -12)]\right\rbrace+\mathcal{O}\left(\frac{1}{r^3}\right),\nonumber
\end{eqnarray}
\begin{equation}
S_1^{(1)}=V_1^{(1)}=0,
\end{equation}
\begin{equation}
\bar{V}_{1}^{(1)}=\int^{\infty}_{r} \frac{xdx}{x^4-1}\int^{x}_1 dy \left[i\omega \bar{V}_1^{(0)}+2 i \omega y \partial_y \bar{V}_1^{(0)}\right]\xrightarrow[]{ r\rightarrow\infty } \frac{3 i\omega}{4 r^2}(\pi+2\log2) +\mathcal{O} \left(\frac{1}{r^3}\right),
\end{equation}
\begin{equation}
\bar{S}_{1}^{(1)}=-\int^{\infty}_{r} \frac{dx}{x^3}\int^{\infty}_x dy\left[y \partial_y \bar{V}_1^{(1)}+12 \partial_y g_4^{(1)}\right]\xrightarrow[]{ r\rightarrow\infty } \mathcal{O}\left(\frac{1}{r^3}\right),
\end{equation}
\begin{equation}
\begin{split}
V_{3}^{(1)}&=-\int^{\infty}_{r} \frac{xdx}{x^4-1}\int^{x}_1 dy \left[i\omega V_3^{(0)}+2i\omega y \partial_y V_3^{(0)}+\frac{12\kappa}{y^3}\bar{\rho}\bar{V}_1^{(1)}-12i \omega \kappa \bar{\rho}  g_4^{(0)} \partial_y f_2 \right]\\
&\xrightarrow[]{ r\rightarrow\infty } \frac{3i\omega \kappa\bar{\rho}}{4r^2} \left[\pi(6-2\pi-3\log2)+\log2(12-9\log2)\right]+\mathcal{O}\left(\frac{1}{r^3}\right),
\end{split}
\end{equation}
\begin{equation}
\begin{split}
\bar{V}_{3}^{(1)}&=-\int^{\infty}_{r} \frac{xdx}{x^4-1}\int^{x}_1 dy \left[i\omega \bar{V}_3^{(0)}+2i\omega y \partial_y \bar{V}_3^{(0)}+\frac{12\kappa}{y^3}\bar{\rho}_5\bar{V}_1^{(1)}-12i \omega \kappa \bar{\rho}_5  g_4^{(0)} \partial_y f_2 \right]\\
&\xrightarrow[]{ r\rightarrow\infty } \frac{3i\omega \kappa\bar{\rho}_5}{4r^2} \left[\pi(6-2\pi-3\log2)+\log2(12-9\log2)\right]+\mathcal{O}\left(\frac{1}{r^3}\right),
\end{split}
\end{equation}
\begin{equation}
\begin{split}
V_{1}^{(2)}&=-\int^{\infty}_{r} \frac{xdx}{x^4-1}\int^{x}_1 dy \frac{12 \kappa}{y^3}q^2(\bar{\rho}_5 V_3^{(0)}+\bar{\rho} \bar{V}_3^{(0)})\\
&\xrightarrow[]{ r\rightarrow\infty }\frac{27 q^2 \kappa^2 \bar{\rho}\bar{\rho}_5} {r^2}[6+\log 2(5\log 2 -12)]+\mathcal{O}\left(\frac{1}{r^3}\right),
\end{split}
\end{equation}
\begin{equation}
\begin{split}
\bar{V}_{1}^{(2)}&=-\int^{\infty}_{r} \frac{xdx}{x^4-1}\int^{x}_1 dy \left[2 i \omega y  \partial_y \bar{V}_1^{(1)}+i \omega \bar{V}_1^{(1)} +\frac{q^2}{y} \bar{V}_1^{(0)} +\frac{12\kappa}{y^3}q^2(\bar{\rho}V_3^{(0)}+\bar{\rho}_5\bar{V}_3^{(0)})-12 \partial_y g_3^{(2)}\right]\\
&\xrightarrow[]{ r\rightarrow\infty }-\frac{1}{16r^2}\left\{\omega^2\left[\pi^2+6(4 \mathcal{C} + (\log2)^2)\right]\right.\\
&\qquad \quad \left. +q^2\left[ 6\pi+\pi^2-12\log2-216\kappa^2(\bar{\rho}^2+\bar{\rho}_5^2)(6+\log 2[5\log 2 -12]) \right]\right\}+\mathcal{O}\left(\frac{1}{r^3}\right),
\end{split}
\end{equation}
\\
sub-sector (ii): $\left\{ S_2,\bar{S}_2, V_4,\bar{V}_4,V_5,\bar{V}_5,V_6,\bar{V}_6 \right\}$:
\begin{equation}
S_2^{(0)}=\bar{S}_2^{(0)}=V_4^{(0)}=\bar{V}_4^{(0)}=V_6^{(0)}=0,
\end{equation}
\begin{equation}
\bar{V}_{6}^{(0)}=-\int^{\infty}_{r} \frac{xdx}{x^4-1}\int^{x}_1 dy 6\left[2 \partial_y g_4^{(0)}+\frac{1}{y^2} \partial_y f_1\right]\xrightarrow[]{ r\rightarrow\infty } - \frac{3 \log 2}{2 r^2}+\mathcal{O}\left(\frac{1}{r^3}\right),
\end{equation}
\begin{equation}
V_{5}^{(0)}=-\int^{\infty}_{r} \frac{xdx}{x^4-1}\int^{x}_1 dy \frac{12 \kappa }{y^3} \bar{\rho} \bar{V}_{6}^{(0)}\xrightarrow[]{ r\rightarrow\infty } \frac{9\kappa \bar{\rho} (\log 2)^2}{4 r^2}+\mathcal{O}\left(\frac{1}{r^3}\right),
\end{equation}
\begin{equation}
\bar{V}_{5}^{(0)}=-\int^{\infty}_{r} \frac{xdx}{x^4-1}\int^{x}_1 dy \frac{12 \kappa } {y^3} \bar{\rho}_5 \bar{V}_{6}^{(0)}\xrightarrow[]{ r\rightarrow\infty } \frac{9\kappa \bar{\rho}_5 (\log 2)^2}{4 r^2}+\mathcal{O}\left(\frac{1}{r^3}\right),
\end{equation}
\begin{equation}
S_2^{(1)}=\bar{S}_2^{(1)}=V_4^{(1)}=\bar{V}_4^{(1)}=V_6^{(1)}=0,
\end{equation}
\begin{equation}
\begin{split}
\bar{V}_{6}^{(1)}&=-\int^{\infty}_{r} \frac{xdx}{x^4-1}\int^{x}_1 dy 6\left[i\omega \bar{V}_6^{(0)}+2i\omega y\partial_y \bar{V}_6^{(0)}+12\partial_y g_4^{(1)}-12i\omega g_4^{(0)}\partial_y f_1 \right]\\
&\xrightarrow[]{ r\rightarrow\infty } -\frac{i\omega}{64r^2}(48 \mathcal{C}+5\pi^2) +\mathcal{O} \left(\frac{1}{r^3}\right),
\end{split}
\end{equation}
\begin{equation}
V_{4}^{(2)}=-\int^{\infty}_{r} \frac{xdx}{x^4-1}\int^{x}_1 dy \frac{12 \kappa }{y^3}q^2 \bar{\rho} \bar{V}_{6}^{(0)}\xrightarrow[]{ r\rightarrow\infty } \frac{9q^2\kappa \bar{\rho} (\log 2)^2}{4 r^2}+\mathcal{O}\left(\frac{1}{r^3}\right),
\end{equation}
\begin{equation}\label{pertf}
\bar{V}_{4}^{(2)}=-\int^{\infty}_{r} \frac{xdx}{x^4-1}\int^{x}_1 dy \frac{12 \kappa } {y^3} q^2\bar{\rho}_5 \bar{V}_{6}^{(0)}\xrightarrow[]{ r\rightarrow\infty } \frac{9q^2\kappa \bar{\rho}_5 (\log 2)^2}{4 r^2}+\mathcal{O} \left(\frac{1}{r^3}\right),
\end{equation}
where $\mathcal{C}$ is the Catalan constant. It is straightforward to read off the boundary data $v_i$ and $\bar{v}_i$ from the solutions presented above.

\section*{Acknowledgements}

YB would like to thank the hospitality of  Department of Physics of Ben-Gurion University of the Negev where this work was initialised and finalised. YB was supported by the Fundamental Research Funds for the Central Universities under grant No.122050205032 and the Natural Science Foundation of China (NSFC) under the grant No.11705037.
TD and ML were supported by the Israeli Science Foundation (ISF) grant \#1635/16 and the BSF grants \#2012124 and \#2014707.

\providecommand{\href}[2]{#2}\begingroup\raggedright\endgroup

\end{document}